\documentclass[fleqn,usenatbib]{mnras}

\usepackage{newtxtext,newtxmath}
\usepackage[T1]{fontenc}

\DeclareRobustCommand{\VAN}[3]{#2}
\let\VANthebibliography\thebibliography
\def\thebibliography{\DeclareRobustCommand{\VAN}[3]{##3}\VANthebibliography}

\usepackage{graphicx}	% Including figure files
\usepackage{amsmath}	% Advanced maths commands

\title[UV photon production of the first stars]{Ultraviolet photon production rates of the first stars: Impact on the He II $\lambda$ 1640 \AA{} emission line from primordial star clusters and the 21-cm signal from cosmic dawn}

\author[J. Wasserman et al.]{Joel Wasserman,$^{1}$
Erik Zackrisson,$^{1}$\thanks{E-mail: erik.zackrisson@physics.uu.se}
Jiten Dhandha,$^{2}$
Anastasia Fialkov,$^{2}$
Leon Noble,$^{3}$
\newauthor
Suman Majumdar$^{3}$
\\
$^{1}$Observational Astrophysics, Department of Physics and Astronomy, Uppsala University, Box 516, SE-751 20 Uppsala, Sweden\\
$^{2}$Institute of Astronomy, University of Cambridge, Madingley Road, Cambridge, CB3 0HA, UK\\
$^{3}$Department of Astronomy, Astrophysics \& Space Engineering, Indian Institute of Technology Indore, Indore 453552, India\\
}

\date{Accepted XXX. Received YYY; in original form ZZZ}

\pubyear{2025}

\begin{document}
\label{firstpage}
\pagerange{\pageref{firstpage}--\pageref{lastpage}}
\maketitle

\begin{abstract}
The first stars, the chemically pristine Population III, likely played an important role in heating the intergalactic medium during the epoch of cosmic dawn. The very high effective temperatures ($\sim 10^5$ K) predicted for the most massive Population III stars could also give rise to tell-tale signatures in the emission-line spectra of early star clusters or small galaxies dominated by such stars. Important quantities in modelling their observational signatures include their photon production rates at ultraviolet energies at which photons are able to ionize hydrogen and helium, dissociate molecular hydrogen and cause Lyman-$\alpha$ heating. Here, we model the spectral energy distributions of Population III stars to explore how these key quantities are affected by the initial mass and rotation of Population III stars given a wide range of models for the evolution of these stars. Our results indicate that rotating Population III stars that evolve to effective temperatures $\sim 2\times 10^5$ K could potentially give rise to a very strong HeII 1640 \AA{} emission line in the spectra from primordial star clusters, without requiring stellar masses of $\gtrsim 100\ \mathrm{M}_\odot$ indicated by previous models for non-rotating Population III stars. The observable impact on 21-cm signatures from cosmic dawn and the epoch of reionization from our set of rotating stars that evolve to $\sim 2\times 10^5$ K is modest, except in case of high Population~III star formation efficiencies which imprint potentially detectable features in the global 21-cm signal and 21-cm power spectrum.
\end{abstract}


%%%%%%%%%%%%%%%%%%%%%%%%%%%%%%%%%%%%%%%%%%%%%%%%%%

%%%%%%%%%%%%%%%%% BODY OF PAPER %%%%%%%%%%%%%%%%%%

\section{Introduction}
The first generation of stars, the chemically pristine Population III (hereafter Pop III), is believed to have played an important role in heating the early intergalactic medium during the epoch of cosmic dawn, in producing the first elements beyond lithium, and in the production of early black holes \citep[for a recent review, see][]{Klessen23}. Such stars probably started forming around redshift $z\approx 30$, about 100 Myr after the Big Bang, but may have continued to form in pockets of primordial gas throughout the reionization epoch and well into the post-reionization era \citep[e.g.][]{Johnson10,Mebane18,Liu20}. In terms of observational probes, 21-cm cosmology may be able to probe the nature of Pop III stars at $z\approx 15$--30 through their impact on the temperature state of the intergalactic medium \citep[e.g.][]{Pochinda:2024, Gessey-Jones25,Ventura25}. At somewhat lower redshifts ($z\approx 6$--15) one may hope to detect Pop III signatures in individual objects, either among samples of gravitationally lensed, high-redshift stars \citep{Windhorst18,Zackrisson24}, in gravitationally lensed Pop III star clusters and galaxies \citep{Zackrisson12,Vanzella23,Fujimoto25}, or within galaxies where Pop III star formation is able to continue alongside metal-enriched star formation \citep[e.g.][]{Sarmento18,Venditti23,Maiolino24,Wang24}. Other ways of empirically studying the properties of Pop III stars include the chemical signatures of Pop III supernovae in gas absorption systems \citep[e.g.][]{Dodorico23,Sodini24,Vanni24}, in high-redshift quasars \citep{Yoshii22} and in second-generation stars in the local Universe \citep[e.g.][]{Fraser17,Ishigaki18,Jiang24,Bonifacio25}, as well as the direct detection of supernovae \citep[e.g.][]{Moriya22,Venditti24}, gamma-ray bursts \citep[e.g.][]{Lazar22}, and tidal disruption events \citep{KarChowdhury24} produced by Pop III stars. 

Due to their chemically pristine composition, Pop III stars are expected to be more massive \citep[e.g.][]{Klessen23}, and reach higher main-sequence temperatures \citep[e.g.][]{Schaerer02} than their metal-enriched descendants. This gives rise to higher hydrogen- and helium-ionizing photon production rates compared to metal-enriched stars, and also differences in the Lyman--Werner (LW; $\mathrm{H}_2$-dissociating) and the Lyman-band rates important for the Wouthuysen--Field coupling during cosmic dawn. However, the production of photons in these different energy regimes depends not only on initial mass but also on stellar rotation \citep[e.g.][]{Yoon12,Murphy21b,Sibony22,Liu25,Liu24,Hawcroft25}, and simulations have indicated that at least some Pop III may rotate at close to break-up speed \citep{Hirano18}. 

Here, we use the Muspelheim \citep{Zackrisson24} models for the spectral energy distributions (SEDs) of Pop III stars to explore the impact of different sets of stellar evolutionary tracks on the ultraviolet (UV) radiation output of Pop III stars. These models use stellar atmosphere SEDs instead of the blackbody SEDs adopted in many previous studies \citep[e.g.][]{Yoon12,Murphy21b,Klessen23}. In section~\ref{sec:models}, we present these models and in section~\ref{sec:fluxes} we report the resulting hydrogen- and helium-ionizing photon production rates, the LW and the Ly-band rates, along with a comparison to results derived from blackbody SEDs. The impact on studies of the UV SEDs of Population III star clusters and galaxies is explored in Section~\ref{sec:HeII1640} and the impact on 21 cm-cosmology is shown in Section~\ref{sec:21cm}. Various future avenues for improvement of the models are discussed in section~\ref{sec:discussion}. Section~\ref{sec:conclusion} summarises our findings.

\section{SED models of Population III stars}
\label{sec:models}
The SED models for Pop III stars (Muspelheim v.1) presented by \citet{Zackrisson24} couple sets of stellar evolutionary tracks to a grid of stellar atmospheres to produce SEDs as a function of Zero Age Main Sequence (ZAMS) mass $M_\mathrm{ZAMS}$ and age. Important simplifications used in the Muspelheim v.1 models include the use of stellar atmospheres for spherical stars without mass loss, and with fixed primordial abundances of hydrogen and helium. Hence, these models neglect the non-sphericity that come from high rotation rates, the effects of winds, and the effects of surface helium enhancement and surface self-pollution of metals during the course of evolution. These shortcomings are further discussed in Section~\ref{sec:discussion}. In this paper, we make use of Muspelheim SED models based on selected Pop III tracks from \citet{Yoon12} at $M_\mathrm{ZAMS}=10$--500 M$_{\odot}$, \citet{Windhorst18} at $M_\mathrm{ZAMS}=10$--100 M$_{\odot}$, \citet{Murphy21a} at $M_\mathrm{ZAMS}=1.7$--120 M$_{\odot}$, \citet{Volpato23} at $M_\mathrm{ZAMS}=100$--1000 M$_{\odot}$ and \citet{Costa25} at $M_\mathrm{ZAMS}=6$--200 M$_{\odot}$. Out of these sets, the \citet{Yoon12} and \citet{Murphy21a} sets feature models both with and without rotation. Although \citet{Yoon12} present stellar evolutionary tracks for several different initial velocities in the range 0.0--0.6 $v_\mathrm{K}$ (where  $v_\mathrm{K}$ is the Keplerian velocity at the equatorial surface), we here limit the discussion to the $v_\mathrm{K}=0.0$ and 0.4 models, for simplicity. 

While many of these tracks are similar close to the ZAMS, they differ significantly in their predictions of their low-$T_\mathrm{eff}$ extension at late stages of stellar evolution. While \citet{Zackrisson24} argue that late evolution to $T_\mathrm{eff}\lesssim 15000$ K have important consequences for the prospects of detecting lensed, individual Pop III stars at high redshifts, one could expect that such differences would be minor when it comes to the UV photon production rates of Pop III stars, since these are dominated by the more long-lived high-$T_\mathrm{eff}$ stages of these stars. However, the \citet{Yoon12} tracks for rotating stars, which also include the effects of magnetic fields, differ from many other models of rotating Pop III stars in the literature \citep[e.g.][]{Murphy21a,Sibony22,Martinet23,Volpato24,Tsiatsiou24} in that some of the stars with high initial rotation velocity evolve to very high $T_\mathrm{eff}$ ($\sim 250 000$ K) at bolometric luminosities an order of magnitude above their ZAMS luminosities at the end of their lifetimes \citep[similar effects are, however, seen in the rotating Pop III models in the recent paper by][]{Hassan25}, with significant impact on their UV properties.

In this paper, we explore the UV properties of Pop III stars in terms of their production of H-, He- and He$^{+}$-ionizing photons, as well as their production of LW and Lyman-band photons. 

The instantaneous photon production rates $Q$, expressed in units of photons per second, are defined as:
\begin{equation}
Q_i = \int_{\lambda_1}^{\lambda_2} \frac{\lambda L_\lambda }{hc} \mathrm{d}\lambda,
\label{eq:Q equation}
\end{equation}
where $L_\lambda$ is the monochromatic luminosity (erg$^{-1}$  s$^{-1}$ \AA$^{-1}$) at wavelength $\lambda$, $h$ is Planck's constant and $c$ the speed of light. The integration limits for the different rates are: $\lambda_1=0$ and $\lambda_2\approx 912$ \AA{} for $Q_\mathrm{H}$; $\lambda_1=0$  and $\lambda_2\approx 504$  \AA{} for $Q_\mathrm{He}$; $\lambda_1=0$ and $\lambda_2\approx 228$  \AA{} for $Q_\mathrm{He^{+}}$; $\lambda_1\approx 912$  \AA{} and $\lambda_2=1107$ \AA{} for $Q_\mathrm{LW}$; $\lambda_1\approx 912$  \AA{} and $\lambda_2=1216$ \AA{} for $Q_\mathrm{Ly}$.
We also present $\epsilon_\mathrm{b}^\mathrm{Ly}$, the Lyman-band photon production count per baryon, integrated over the stellar lifetime $t_\mathrm{lifetime}$, given by:
\begin{equation}
\epsilon_\mathrm{b}^\mathrm{Ly} = \frac{m_\mathrm{p}}{M_\mathrm{ZAMS}} \int_0^{t_\mathrm{lifetime}} Q_\mathrm{Ly}(t) \mathrm{d}t,
\end{equation}
where $m_\mathrm{p}$ is the mass of a proton.

While the procedure for generating stellar SEDs is the same as in \citet{Zackrisson24}, the numerical computation of $Q$s has been improved, and the set of rotating stars presented here is also wider in terms of $M_\mathrm{ZAMS}$ than that released in connection to the \citet{Zackrisson24} paper.

\section{Ionizing and Lyman--Werner fluxes}
\label{sec:fluxes}
\begin{figure}
\includegraphics[scale=0.4]{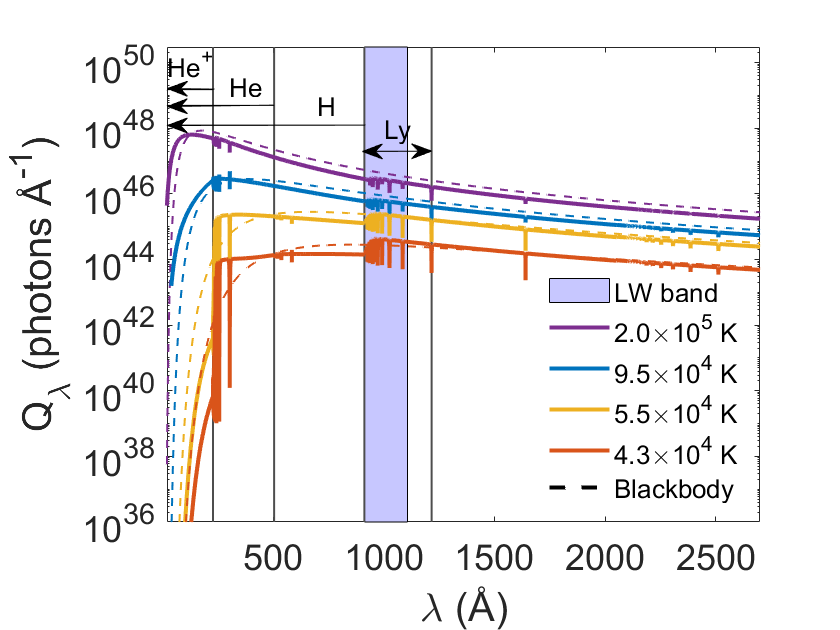}
\caption{SEDs (in units of photon production rate per unit wavelength) of TLUSTY stellar atmosphere models with primordial chemical composition for  $T_\mathrm{eff}= 2.05\times 10^5$ K (purple solid line),
$T_\mathrm{eff} = 9.5\times 10^4$ K (blue solid line),
$T_\mathrm{eff} = 5.5\times 10^4$ K (yellow solid line) and $T_\mathrm{eff} = 4.3\times 10^4$ K (red solid line), contrasted to the corresponding blackbody SEDs (purple, blue, yellow and red dashed lines). To avoid overlap, the SEDs are scaled to bolometric luminosities $\log(L_\mathrm{bol}/L_\odot)\approx 6.8$, 5.3, 4.3  and 3.3, respectively. The surface gravities of the TLUSTY SEDs has been set to $\log(g)=6.5$ for the $T_\mathrm{eff}= 2.05\times 10^5$ K model and 5.0 in the other cases. The coloured patch indicates the extent of the LW band and the arrows the extent of the Lyman band and the H-, He-, He$^{+}$-ionizing part of the SEDs. Differences between the stellar atmosphere SEDs and the blackbody SEDs are seen in all of the different bands, although the degree to which blackbody models over- or underpredict the rates depend on $T_\mathrm{eff}$. 
The $T_\mathrm{eff}=4.3\times 10^4$, $5.5\times 10^4$ and $9.5\times 10^4$ K SEDs are considered representative of early evolution along the $9 \ \mathrm{M}_\odot$, $15 \ \mathrm{M}_\odot$  and $120 \ \mathrm{M}_\odot$ \citet{Murphy21a} Pop III tracks for both rotating and non-rotating stars, whereas the $\approx 2\times 10^5$ K model is representative only for rotating \citet{Yoon12} 20--150 M$_{\odot}$ stars at the very end of their lifetimes.}
\label{fig:atmos vs BB}
\end{figure}

As previously shown by e.g. \citet{Bromm01}, \citet{Larkin23}, and \citet{Liu25}, the SEDs of stellar atmosphere models predict UV properties of hot Pop III stars that deviate from those of blackbody SEDs scaled to the same effective temperatures $T_\mathrm{eff}$ and bolometric luminosities $L_\mathrm{bol}$. We illustrate this in Fig. \ref{fig:atmos vs BB}, where we compare the SEDs of the TLUSTY stellar atmosphere models used in Muspelheim to those of blackbody SEDs for stars at $T_\mathrm{eff} \approx 4\times 10^4$--$2\times 10^5$ K. The photon counts per wavelength ($Q_\lambda$) differ between blackbody and stellar atmosphere SEDs in {\it all} the UV bands, albeit to different extent and in different directions throughout the range of model SEDs shown. It is worth noting, that even though the highest-$T_\mathrm{eff}$ SED shown ($\approx 2\times 10^5$ K, a temperature reached only by rotating \citet{Yoon12} 20--150 M$_{\odot}$ Pop III stars in our set) exhibits the most blackbody-like shape (with small continuum breaks only), the peak of the stellar-atmosphere SED occurs at shorter wavelengths compared to the corresponding blackbody, an effect previously seen in the stellar atmosphere SEDs for Pop III stars by \citet{Bromm01} and also for pure hydrogen atmospheres for the very hot central stars of planetary nebulae \citep{Rauch03}. 

\begin{figure*}
    \centering
    \begin{minipage}[t]{0.496\linewidth}
        \centering
        \includegraphics[width=\linewidth]{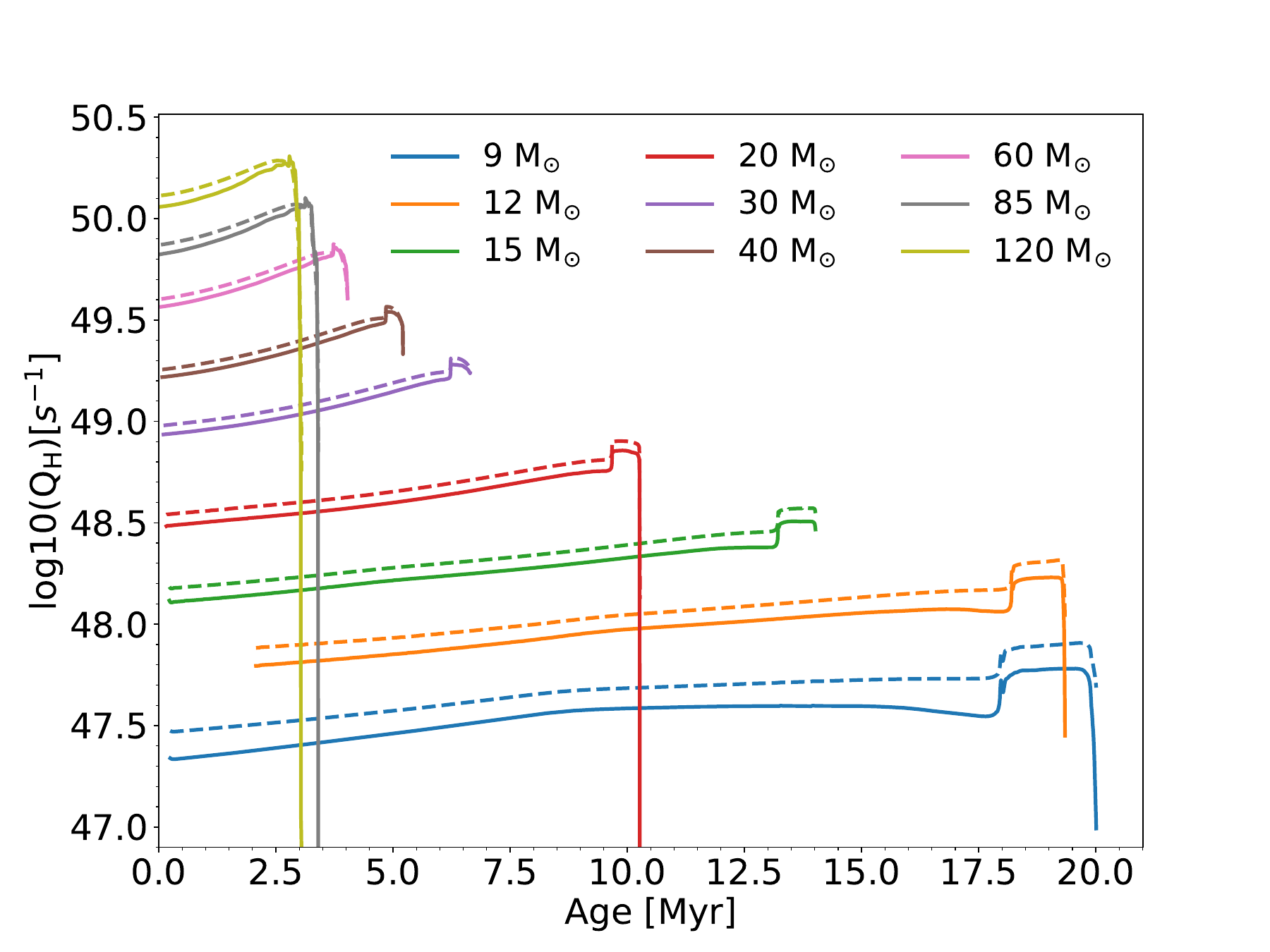}
    \end{minipage}
    \hfill
    \begin{minipage}[t]{0.496\linewidth}
        \centering
        \includegraphics[width=\linewidth]{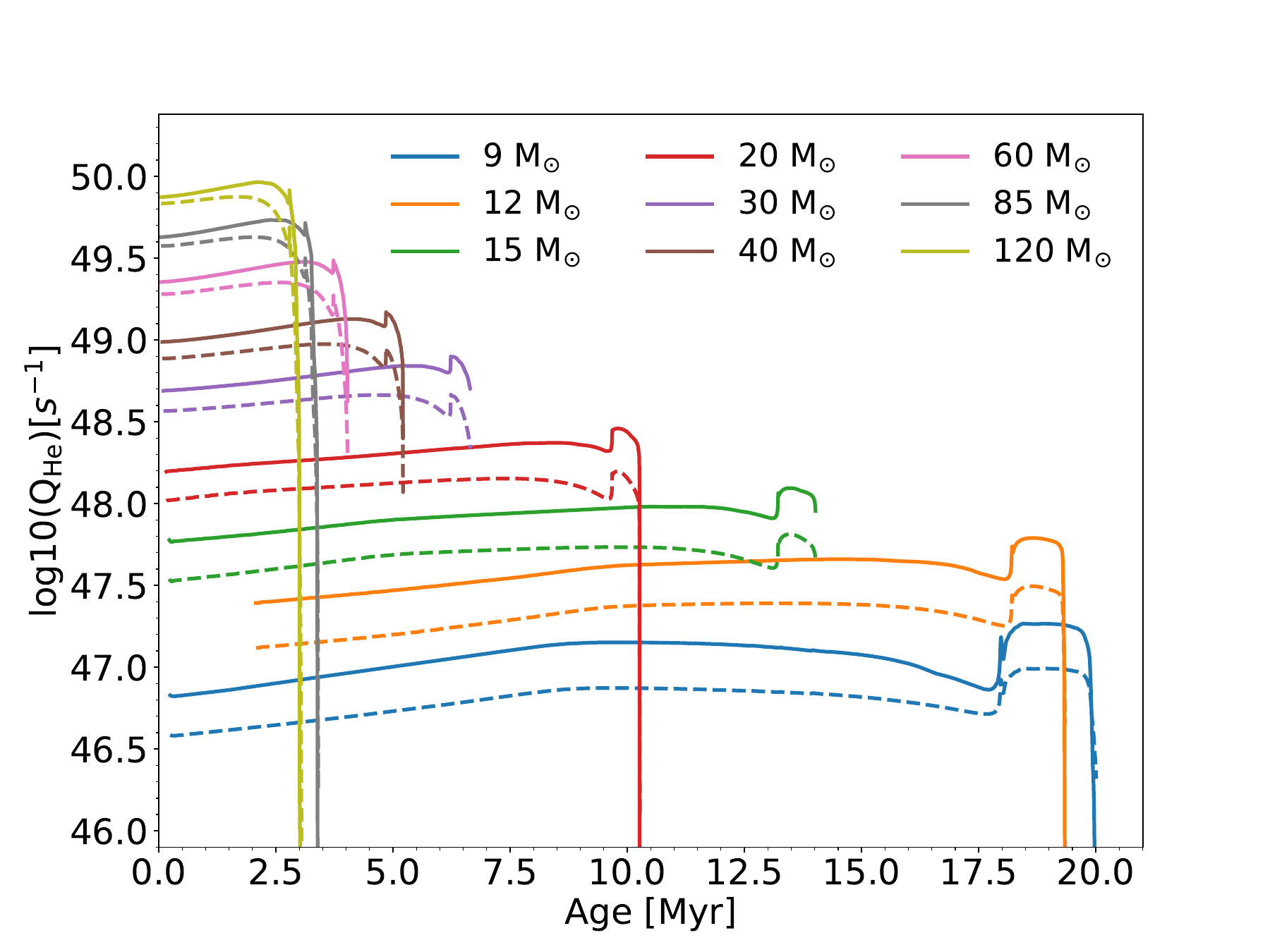}
    \end{minipage}
    \centering
    \begin{minipage}[t]{0.496\linewidth}
        \centering
        \includegraphics[width=\linewidth]{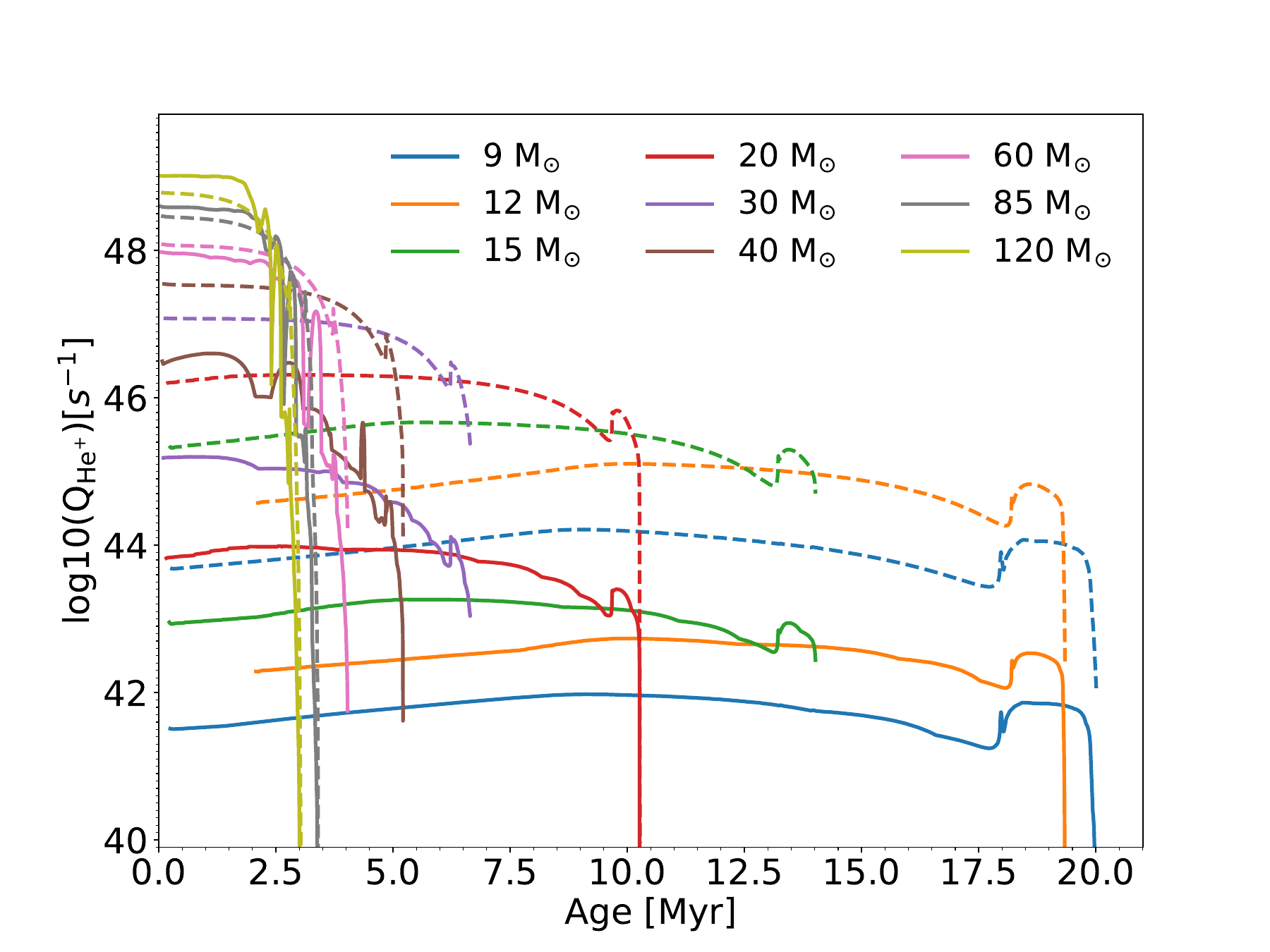}
    \end{minipage}
    \hfill
    \begin{minipage}[t]{0.496\linewidth}
        \centering
        \includegraphics[width=\linewidth]{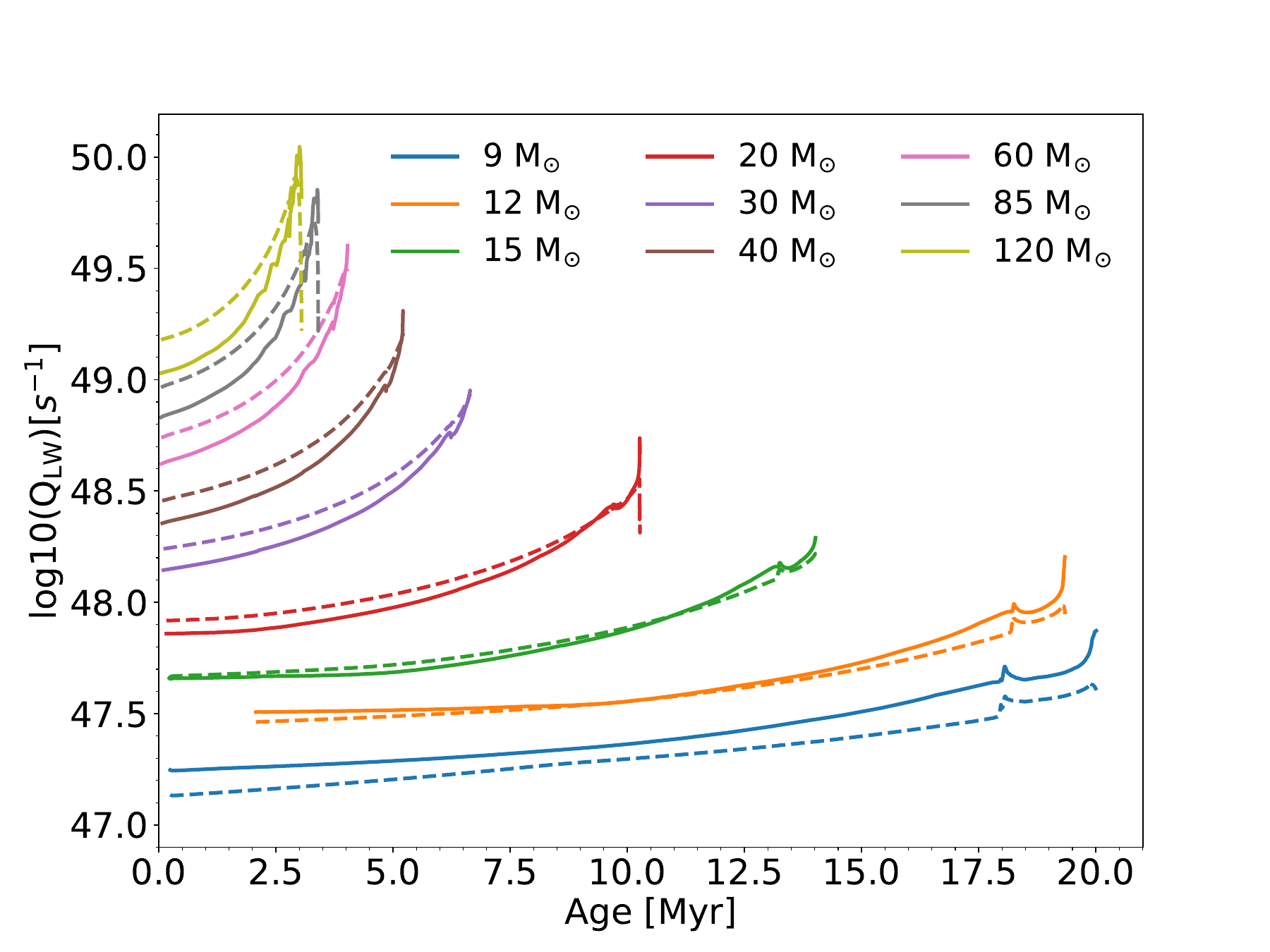}
    \end{minipage}
    \caption{Photon productions rates ($Q_i$) as a function of stellar age for \citet{Murphy21b} non-rotating models. The top left panel is for H-ionizing photons, the top right He-ionizing, the bottom left He$^+$-ionizing, and the bottom right LW-photons. Solid lines represent predictions based on stellar atmospheres and dashed lines the corresponding predictions for black body spectra. The largest difference between photon production rates of black body and atmospheres is the over-prediction of He$^{+}$-ionizing photons for black bodies, reaching up to $\sim2.3$ dex for the less massive stars. As expected from Fig. \ref{fig:atmos vs BB}, the effect becomes smaller for the higher-mass stars and even inverses for the most massive, as a result of their high $T_\mathrm{eff}$ ($\approx 9.5\times 10^4$ K for the 120 M$_{\odot}$ model) close to the ZAMS. The other photon production rates are less drastically affected.}
    \label{fig:bbcomp}
\end{figure*}

The effect on the ionizing- and LW photon production rates can be seen in Fig. \ref{fig:bbcomp} where we compare results generated from blackbody SEDs (dashed lines; in good agreement with the results presented by \citealt{Murphy21b}) and stellar atmospheres (solid lines) for the non-rotating \citet{Murphy21a} stellar evolutionary tracks with initial masses in the range 9--120 M$_{\odot}$. The general behaviour follows what we would expect from Fig. \ref{fig:atmos vs BB}, as the shape\footnote{But not the $L_\mathrm{bol}$ scaling, which in Fig. \ref{fig:atmos vs BB} has been chosen to avoid overlap of the SEDs in the plot} of the $T_\mathrm{eff}=4.3\times 10^4$ K, $T_\mathrm{eff}=5.5\times 10^4$ and $T_\mathrm{eff}=9.5\times 10^4$ SEDs in Fig. \ref{fig:atmos vs BB} are representative of early evolution along the $9 \ \mathrm{M}_\odot$, $15 \ \mathrm{M}_\odot$  and $120 \ \mathrm{M}_\odot$ \citet{Murphy21a} tracks in Fig. \ref{fig:bbcomp}. The blackbody SEDs have consistently higher H-ionizing rates, but lower He-ionizing rates, for all the stellar masses plotted, although by no more than $\approx0.2$-0.3 dex. The behaviour of the He$^+$-ionizing rates is more complex, as the blackbody under-predicts the flux at 85-120 M$_{\odot}$ by $\approx 0.1$-0.3 dex, but then severely starts to over-predict the He$^+$-ionizing flux by up to $\approx 2$ dex at lower masses due to the presence of a strong continuum break in the stellar atmosphere SEDs at the He$^+$ edge (seen in the $T_\mathrm{eff} = 5.5\times 10^4$ K and $4.3\times 10^4$ K SEDs in Fig. \ref{fig:atmos vs BB}). In the case of the LW flux, the blackbody SED over-predicts the flux by up to $\approx 0.15$ dex at the highest masses (as seen in the $T_\mathrm{eff} = 9.5\times 10^4$ K SED), becomes nearly identical at 15 M$_{\odot}$ (as seen in the $T_\mathrm{eff} = 5.3\times 10^4$ K SED) and drops below the stellar atmosphere SED by $\approx 0.1$ dex by 9 M$_{\odot}$ (as seen in the $T_\mathrm{eff} = 4.3\times 10^4$ K SED).

\begin{figure*}
    \centering
    \begin{minipage}[t]{0.49\linewidth}
        \centering
        \includegraphics[width=\linewidth]{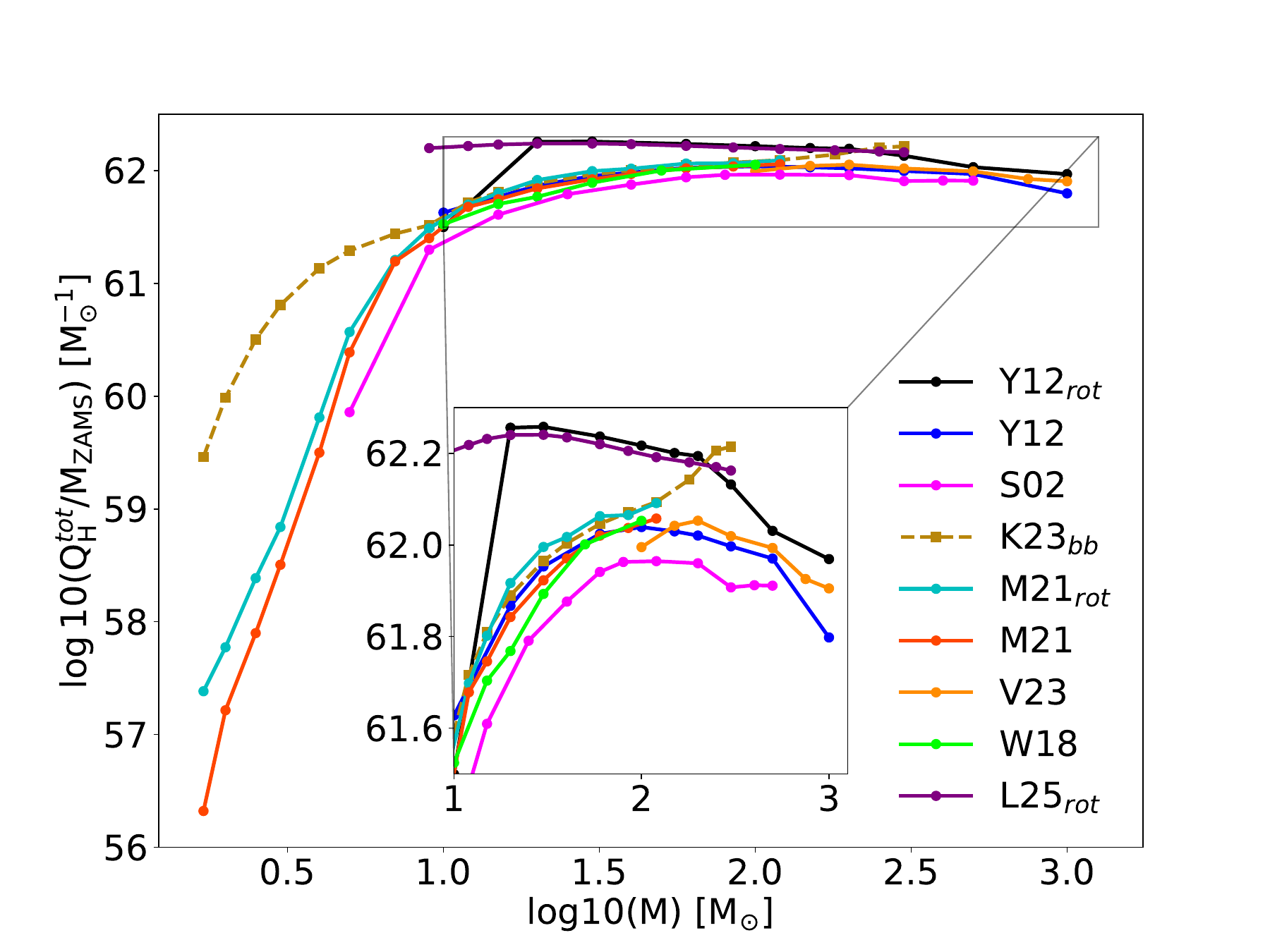}
    \end{minipage}
    \hfill
    \begin{minipage}[t]{0.49\linewidth}
        \centering
        \includegraphics[width=\linewidth]{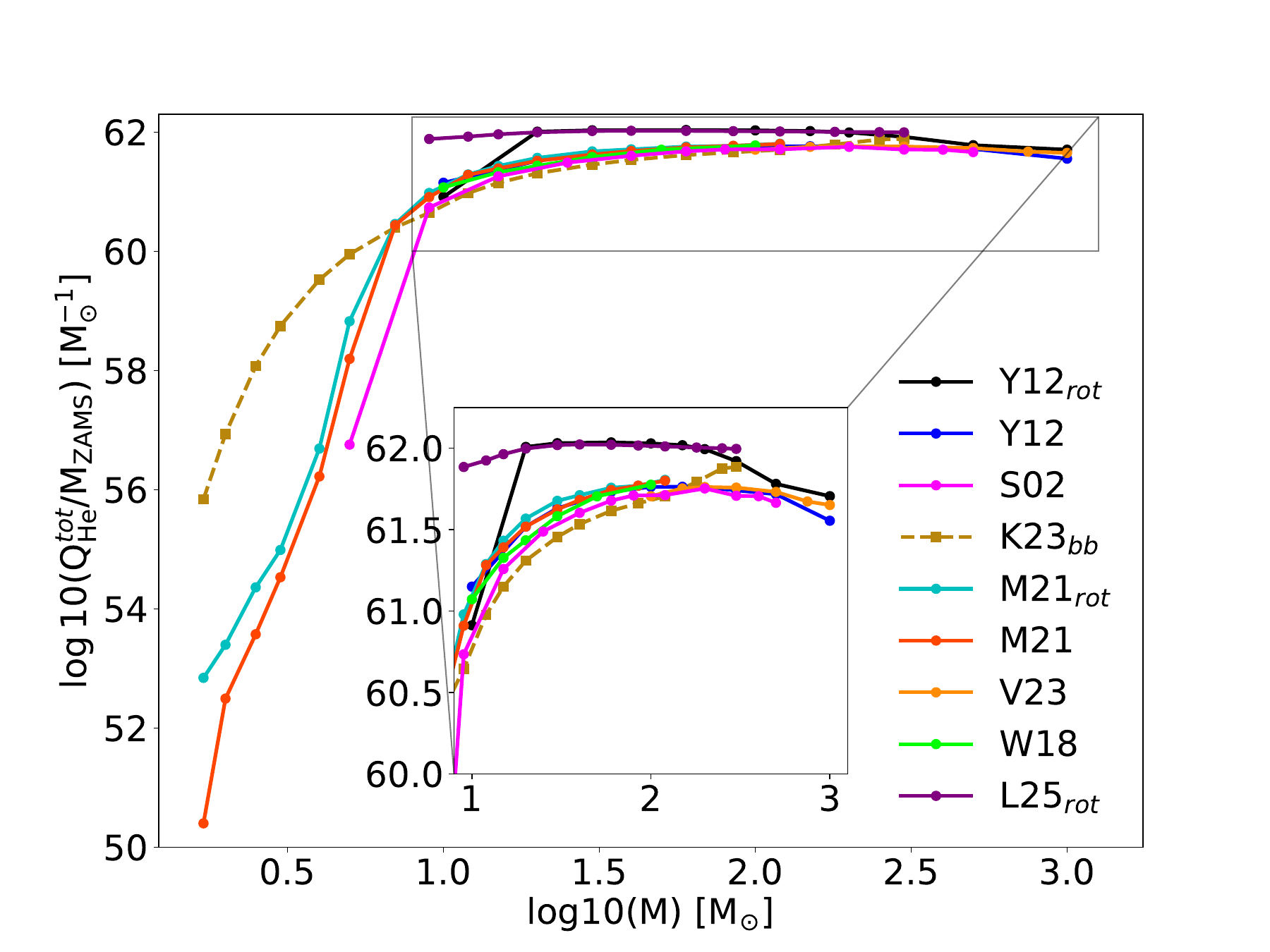}
    \end{minipage}
    \centering
    \begin{minipage}[t]{0.49\linewidth}
        \centering
        \includegraphics[width=\linewidth]{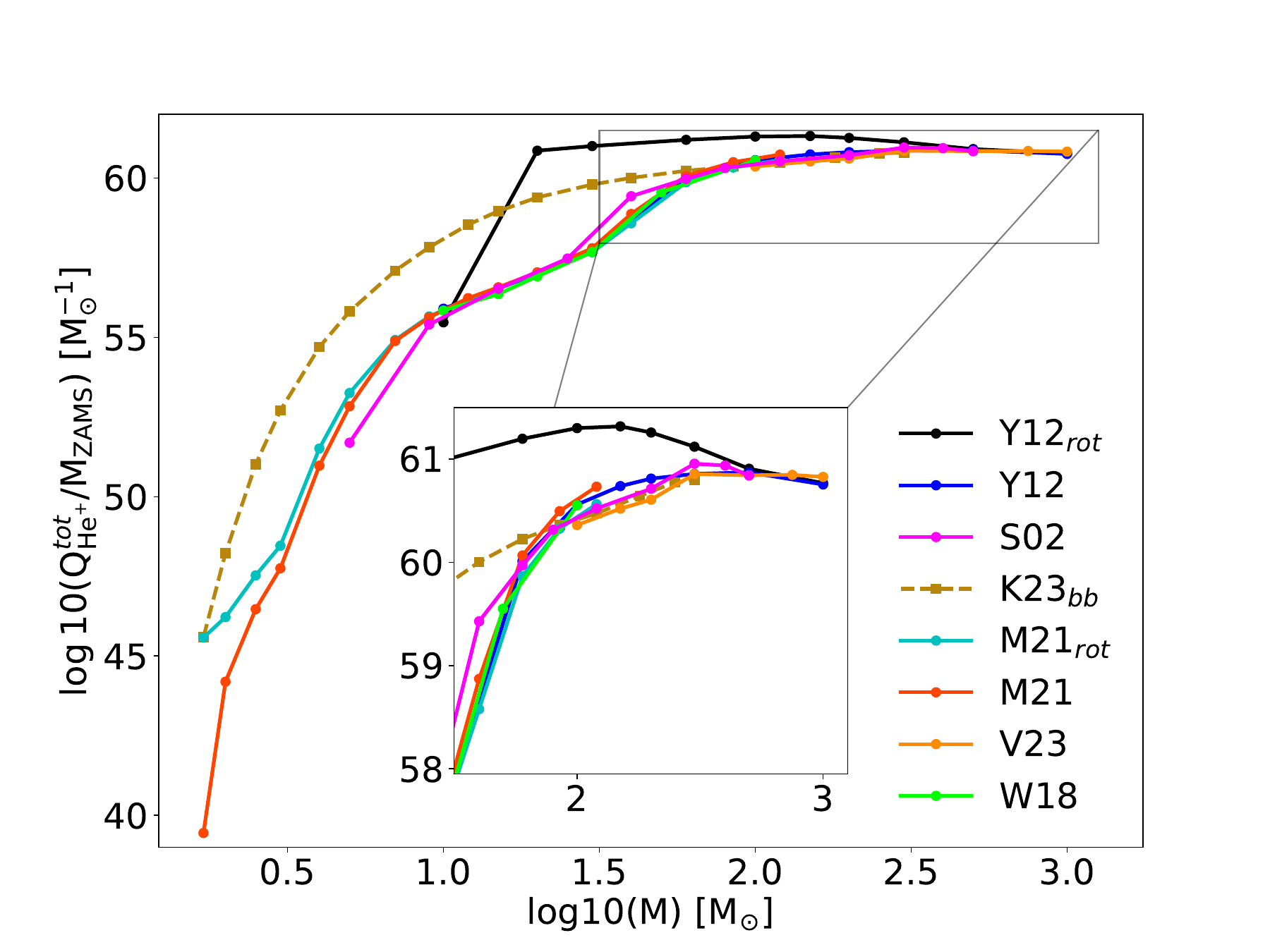}
    \end{minipage}
    \hfill
    \begin{minipage}[t]{0.49\linewidth}
        \centering
        \includegraphics[width=\linewidth]{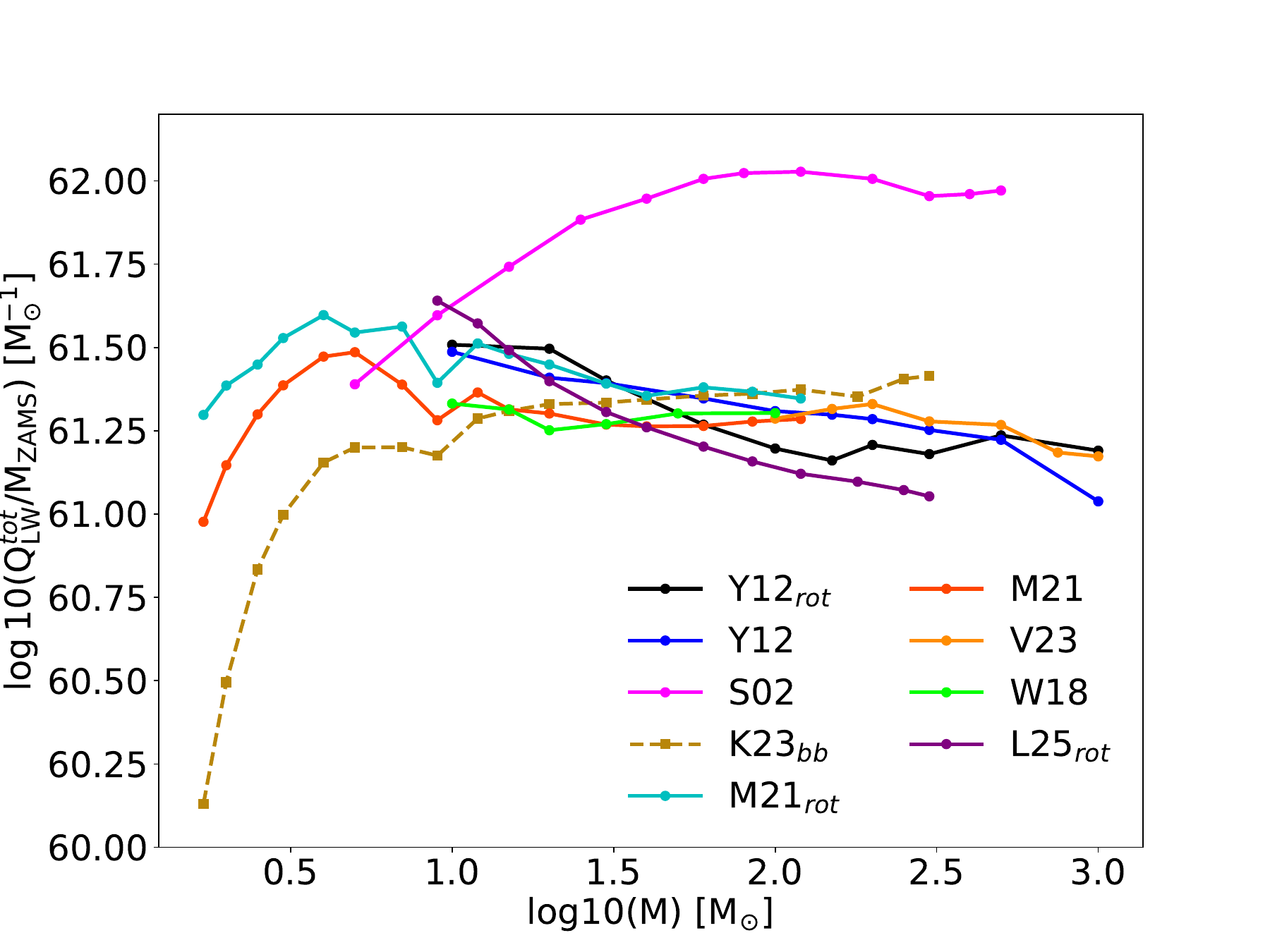}
    \end{minipage}
    \caption{Lifetime-integrated UV photon productions rates per solar mass from rotating and non-rotating models with tracks from \citet[][black and blue; Y12$_\mathrm{rot}$ and Y12]{Yoon12}, \citet[][teal and red; M21$_\mathrm{rot}$ and M21]{Murphy21b}, \citet[][yellow; V23]{Volpato23}, \citet[][green, W18]{Windhorst18}, \citet[][maroon; K23$_\mathrm{bb}$]{Klessen23}, \citet[][magenta; S02]{Schaerer02}, and \citet[][purple; L25$_\mathrm{rot}$]{Liu25}. 
    The top left gives the values for H-ionizing photons, the top right for He-ionizing, the bottom left for He$^{+}$-ionizing, and the bottom right for LW photons. The values from \citet{Klessen23} are based on tracks from \citet{Murphy21b} and \citet{Martinet23}, using black body spectra which is why they differ from the other models, especially at lower masses. \citet{Liu25} is omitted from the He$^+$ panel as they only include wavelengths above 124\AA{}. In general, all models based on stellar atmosphere SEDs agree on the H-ionzing and He-ionizing fluxes to within $\approx$ 0.3 dex for non-rotating stars at $M\gtrsim$ 10 M$_{\odot}$ and within $\approx$ 0.7 dex when rotation is included for these masses. The use of black body spectra severely overpredict the He$^+$-ionizing fluxes compared to comparable models based on stellar atmosphere SEDs. However, models for rotating \citet{Yoon12} stars at 20--300 M$_{\odot}$ produce higher He$^+$-ionizing fluxes than all other models. In the LW-band, the \citet{Schaerer02} predictions constitute a significant outlier which we discuss further in section \ref{subsec: LW}.} 
    \label{fig:Q_lifetime_total}
\end{figure*}

When modelling the impact of Pop III stars on the intergalactic medium, the detailed temporal evolution of the UV photon rates is often neglected in favour of lifetime-integrated quantities \citep[for a counterexample, see][]{Gessey-Jones22}. This may be reasonable in cases where the lifetimes of the Pop III stars are short compared to timescales related to the thermal evolution of the IGM or changes in cosmic star formation rate density, as is the case for the more massive Pop III stars. In Fig. \ref{fig:Q_lifetime_total} we present the lifetime-integrated photon production rates of our Pop III models per solar mass, compared to blackbody models from \citet{Klessen23} and models for rotating stars with homogeneous chemical evolution from \citet[][further discussed in section~\ref{sec:21cm}]{Liu25}. For stellar masses $M\gtrsim 10\ \mathrm{M}_\odot$, all models except the \citet{Yoon12} and \citet{Liu25} rotating models agree on the H-ionizing and He-ionizing rates to within $\approx 0.3$ dex. The \citet{Yoon12} models at 20 $\lesssim$ M$_{\mathrm{ZAMS}}$ $\lesssim$ 200 M$_{\odot}$, and all of the \citet{Liu25} models, undergo chemically homogeneous evolution which significantly boosts the ionizing photon production of these stars. At lower masses, models based on blackbodies are seen to deviate from those based on stellar atmosphere SEDs by up to $\approx 3$ dex for H and $\approx 5$ dex for He.

In the case of the He$^{+}$-ionizing photon production rates, models based on blackbody SEDs at $M\lesssim 100\ \mathrm{M}_{\odot}$ consistently produce higher values than corresponding models based on stellar atmosphere SEDs, as expected from Fig.~\ref{fig:bbcomp}, with an offset that reaches $\approx 6$ dex at the lowest masses. Models for rotating \citet{Yoon12} stars at 20--300 M$_{\odot}$ produce higher He$^+$-ionizing rates than all other models at comparable masses. This increase of He$^+$-ionzing photons due to rotation was already highlighted by \citet{Yoon12}, although the increase compared to non-rotating models becomes much higher in models based on stellar atmosphere SEDs (peaking at $\approx 3.8$ dex at 20 M$_{\odot}$) compared to the blackbody SEDs ($\lesssim 1.4$ dex) used by \citet{Yoon12}, as shown in their fig. 7. \citet{Liu25} is not included in the He$^{+}$ panel as their spectra are cut of at 124\AA{} which leaves out a significant portion of the He$^{+}$-ionizing photons, especially at high T$_{\mathrm{eff}}$ (Fig. \ref{fig:atmos vs BB}).

For the LW band, models that undergo CHE produce somewhat lower rates ($\approx 0.2$--0.3 dex) than other models at $\gtrsim$ 50 M$_{\odot}$ while lower masses produce rates on the higher side compared with other models. The \citet{Klessen23} blackbody models also produce systematically lower rates at low stellar masses (reaching offsets of $\approx 0.85$ dex at 1.7 M$_{\odot}$) compared to models based on stellar atmosphere SEDs. At high masses, the most notable outliers are the models by \citet{Schaerer02}, which at $M>5\ \mathrm{M}_\odot$ produce much higher rates than all other models (up to $\approx 0.7$ dex higher). This discrepancy is discussed further in Section \ref{subsec: LW}.

The lifetime-average photon production rates corresponding to the lifetime-integrated rates in Fig. \ref{fig:Q_lifetime_total} for our \citet{Yoon12} and \citet{Murphy21a} models can be found in Table~\ref{tab:Qs}. $Q(t)$ tables and both lifetime-integrated and lifetime-average photon production rates for all models can be downloaded from the Muspelheim webpage (see Section \ref{sec:data availability}). 

\section{Impact on the the HeII$\lambda$1640 emission line from Population III star clusters}
\label{sec:HeII1640}
The very high He$^{+}$-ionizing photon production rates of Pop III rotating \citet{Yoon12} stars are interesting in the context of HeII emission lines from high-redshift galaxies. In particular, the HeII$\lambda$1640 emission line has been proposed as a potential Pop III signature, as it can become very strong (with rest-frame equivalent widths up to $\mathrm{EW}_{1640}\sim 100$ {\AA}) in the presence of $T_\mathrm{eff} \sim 10^5$ K stars \citep[e.g.][]{Tumlinson00,Schaerer02,Raiter10,Nakajima22}. For non-rotating stars, effective temperatures required to produce
$\mathrm{EW}_{1640}\gtrsim 30$ {\AA}
are not reached until $\sim 100\ \mathrm{M}_\odot$, which means that a very strong HeII$\lambda$1640 line from an integrated stellar population (Pop III star cluster or Pop III-dominated galaxy) could serve as an indicator of a top-heavy Pop III stellar initial mass function (IMF). Recently, a system with $\mathrm{EW}_{1640}\gtrsim 30$ {\AA} and no detectable metal lines, located in the halo of the galaxy GNz-11 at $z\approx 10.6$, was found by \citet{Maiolino24} who interpreted this as evidence of a Pop III IMF extending at least up to 500 M$_{\odot}$ and a total mass of $\sim 2\times 10^5\ M_\odot$ in Pop III stars.

\begin{figure}
    \centering
    \includegraphics[width=\linewidth]{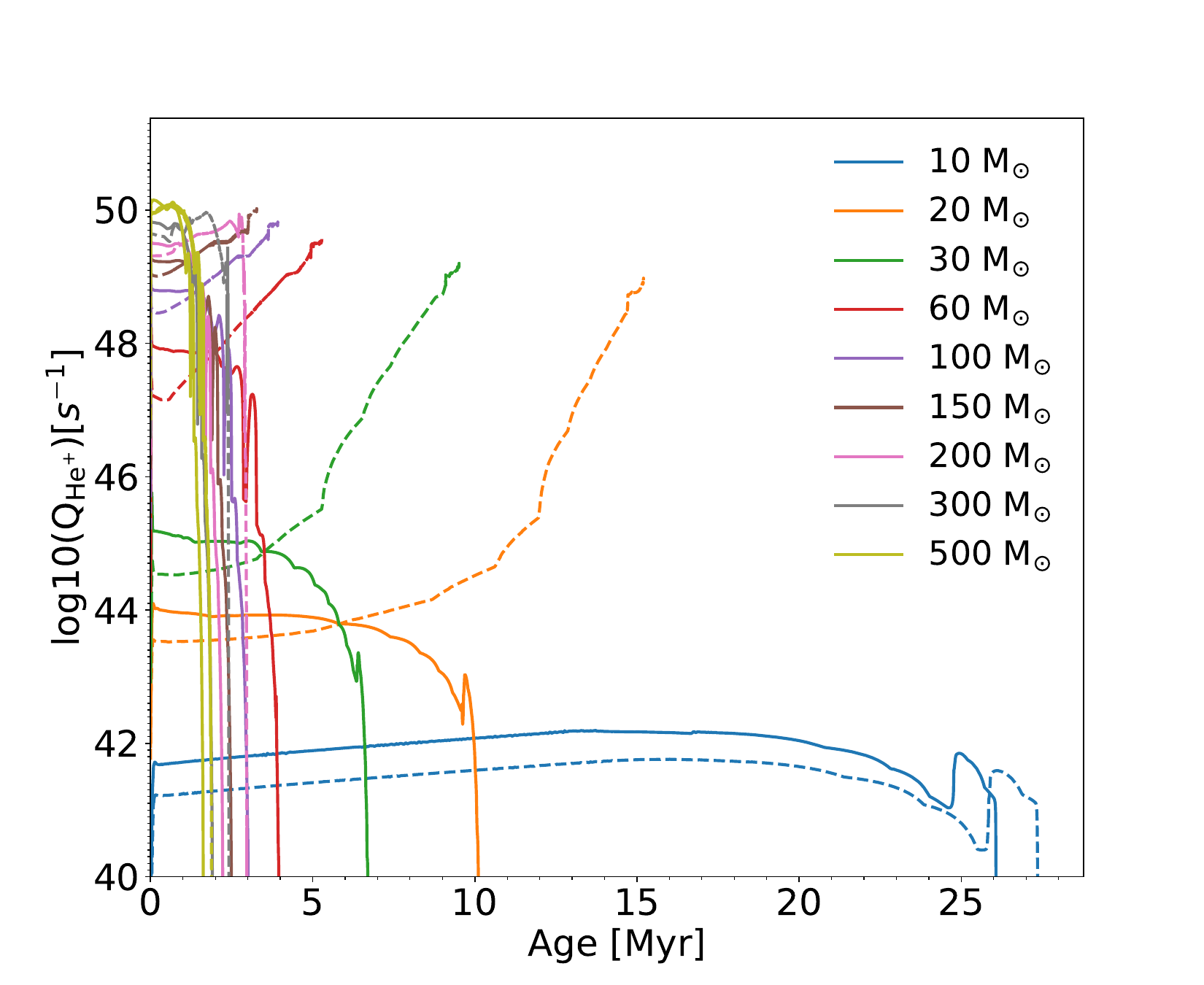}
    \caption{Comparison of He$^{+}$ ionizing fluxes for rotating and non-rotating tracks from \citet{Yoon12}. The dashed lines show rotating models with initial velocity v = 0.4v$_k$ and the solid lines show the corresponding non-rotating models. Large differences between rotating and non-rotating stars are seen for $>10\ \mathrm{M}_{\odot}$ due to extended lifetimes for rotating stars and the very high temperatures ($\approx 2\times 10^5$ K) that 20--150 M$_{\odot}$ rotating stars evolve to at the end of their lifetimes.}
    \label{fig:QHeII_vs_t}
\end{figure}

In Fig. \ref{fig:QHeII_vs_t}, we explore the temporal evolution of the He$^{+}$-ionizing rates of our \citet{Yoon12} models, with and without rotation. Since rotation in these models shifts the early main sequence to lower $T_\mathrm{eff}$ and slightly lower bolometric luminosities, rotation leads to lower He$^+$ rates during early evolution (although this phase is very brief in the case of the 500 M$_{\odot}$ model). However, since the rotating stars eventually evolve to become hotter and more luminous than their non-rotating counterparts for $M>10 \ \mathrm{M}_\odot$, the He$^+$-ionizing photon rates of these rotating stars grow by several orders of magnitude at high ages. This effect, coupled to the extended lifetimes of these stars (due to rotational mixing renewing the fuel in the core), explains the boosted time-integrated $Q_{He^+}$ rates for rotating \citet{Yoon12} models shown in Fig. \ref{fig:Q_lifetime_total}.

Since Fig. \ref{fig:QHeII_vs_t} indicates that the He$^+$-ionizing rates of these 20--30 M$_{\odot}$ rotating models can become comparable, for about $\sim 1$ Myr at the end of their lifetimes, to the peak He$^+$-ionizing rates of 100-150 M$_{\odot}$ stars (maintained over a similar timescale, but during early evolution), it seems that very strong HeII$\lambda$1640 emission in principle could be produced by rotating Pop III stars without the presence of $M\gtrsim 100\ \mathrm{M}_\odot$ stars.

To explore this, we use the latest version of the photoionization code \textsc{cloudy} (v23.01), last described by \citet{Cloudy1} and \citet{Cloudy2}, to investigate the effect of gas parameters on the line strength and equivalent width of the HeII$\lambda$1640 emission line in gas clouds surrounding single stars from the \citet{Yoon12} set. We assume primordial gas abundances, a plane-parallel nebula, and an ionization parameter in the range  $\log U = -1$ to $-4$.

\begin{figure}
    \centering
    \includegraphics[width=\linewidth]{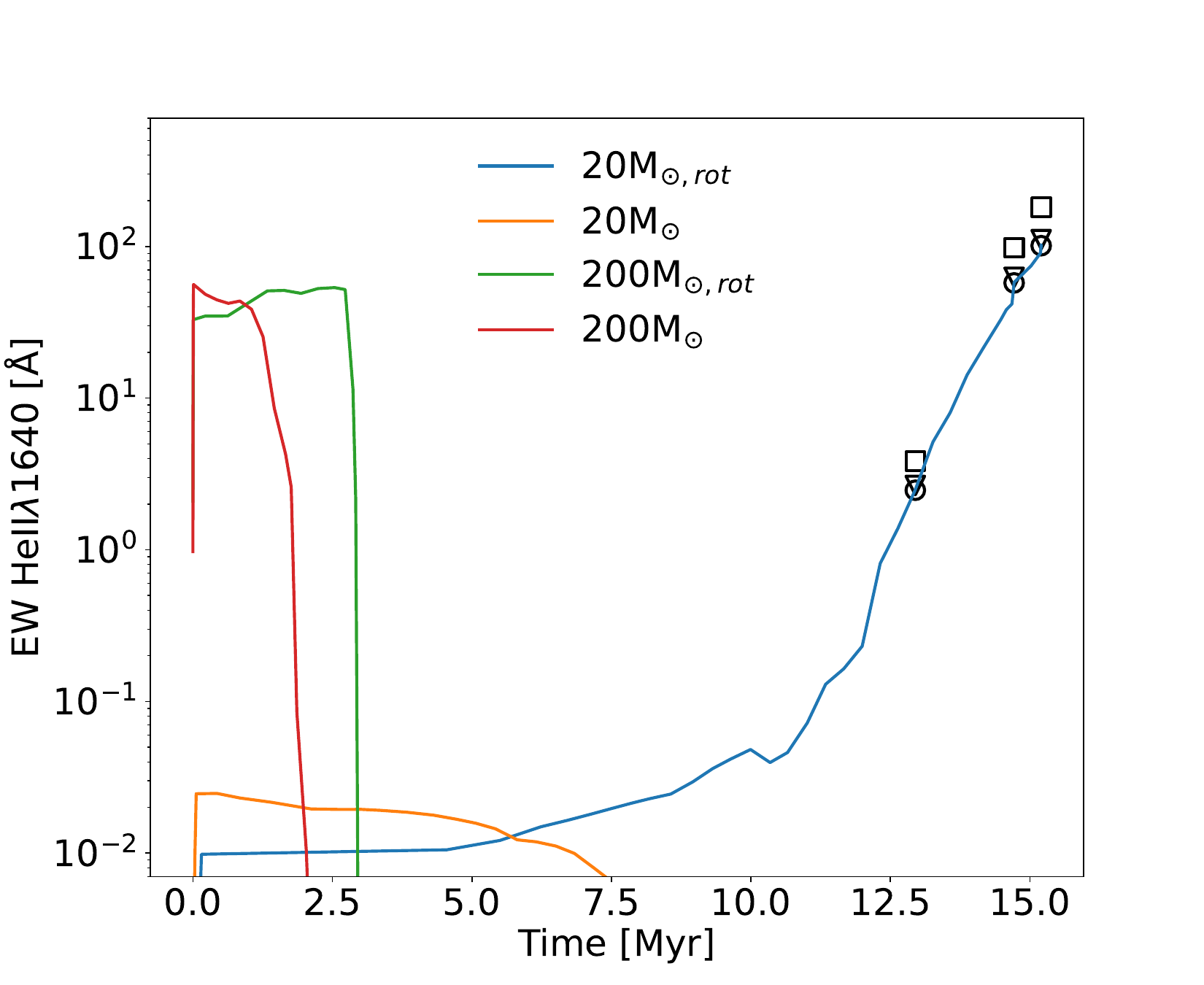}
    \caption{Equivalent widths of the HeII$\lambda$1640 line as a function of time for the \citet{Yoon12} 20 and 200 M$_{\odot}$ rotating and non-rotating stars. The parameters used in the \textsc{cloudy} calculations are log U = -2 and n(H) = 10$^{2}$. At three points we show the equivalent widths for hydrogen density n(H) = 10$^{1}$ (circles), 10$^{3}$ (triangles), and 10$^{4}$ (squares). The 20 M$_{\odot}$ star reaches equivalent widths surpassing the maximum of the 200 M$_{\odot}$ stars at the end of its lifetime. Increasing the hydrogen density in the nebular cloud slightly increases the predicted equivalent width.}
    \label{fig:EW_new}
\end{figure}

In Fig. \ref{fig:EW_new} we show the resulting equivalent widths of the HeII$\lambda$1640 emission line for a subset of  \citet{Yoon12} tracks.  Comparing our results with  HeII$\lambda$1640 EWs from \citet{Schaerer03} and \citet{Raiter10} -- which both assume fully sampled Pop III IMFs -- we are able to obtain only slightly higher HeII$\lambda$1640 equivalent widths than the most extreme values seen in their models. However, the major difference is that both  \citet{Schaerer03} and \citet{Raiter10} require IMFs dominated by very massive stars to reach the high EWs that we are able to obtain for 20 M$_{\odot}$ stars. We also explore other ionization parameters, not shown in the figure, and find that the equivalent widths are largely stable at log(U) $\geq$ -3, but note a significant drop at log(U) = -4 in agreement with \citet{Raiter10}.

Our results indicate that a stellar population of \citet{Yoon12} rotating stars with an IMF peaking at $\sim 20\ \mathrm{M}_\odot$ in principle could produce equally strong HeII$\lambda$1640 equivalent widths as what has been seen for Pop III models with non-rotating stars and more top-heavy IMFs  (like the 50-500 $M_\odot$, Salpeter-slope model, and the $M_c=60$, $\sigma=1$, 1--500 M$_{\odot}$ lognormal model from \citealt{Raiter10}). Invoking rotating \citet{Yoon12} $\sim 20\ \mathrm{M}_\odot$ Pop III stars to explain the HeII$\lambda$1640 emission in the halo of GNz-11 would not only remove the requirement of $M\gtrsim 100\ \mathrm{M}_\odot$ Pop III stars to explain the high equivalent width, but also slightly reduce the total mass in Pop III stars required to explain the line luminosity. Because our rotating $\sim 20\ \mathrm{M}_\odot$ stars reach somewhat higher $Q_\mathrm{He^{+}}/M_\mathrm{ZAMS}$ ratio than the models used by \citet{Maiolino24}, we find that the total mass in Pop III stars could be reduced by a factor of $\sim 2$ compared to their estimate, although this would happen only in a scenario where {\it all} Pop III stars within the system are rapidly rotating $\sim 20\ \mathrm{M}_\odot$ stars caught right in their hottest phase. An IMF strongly peaked around $\sim 20 \mathrm{M}_\odot$ is not favoured by current simulations \citep[for a review, see][]{Klessen23}, but the chemical properties of several extremely metal-poor stars in the Milky Way halo and in ultrafaint dwarfs can be explained by supernovae in the $\lesssim 40 \ M_\odot$ mass range, either from faint supernovae with fallback, hypernovae or a combination of supernovae and winds from rapidly rotating Pop III stars \citep[e.g.][]{Ishigaki18,Skuladottir21,Placco21,Jeena23}. Under the assumption of second-generation stars having been enriched by single Pop III supernovae, the exercise of fitting models of supernova ejecta to stellar abundance patterns can give the appearance of a log-normal Pop III IMF peaked around $\sim 25 \ \mathrm{M}_{\odot}$ \citet{Ishigaki18}, but folding in duo-enrichment and the propensity for failed supernovae can also make observed abundance patterns consistent with an IMF extending to much lower and higher masses \citep{Jiang24}.
Finally, a Pop III IMF peaked around 20 $M_\odot$ is also consistent with the Pop III IMF range found by cosmological models of Milky Way formation, which self-consistently follow the chemical evolution across cosmic time and constrain the Pop III IMF by matching the number and properties of observed ancient metal-poor stars \citep[e.g.][]{Koutsouridou24}.

It is important to note that since the $T_\mathrm{eff}\sim 10^5$ K phase of the 20 M$_{\odot}$ rotating \citet{Yoon12} model is reached only at the very end of its $\approx 15$ Myr lifetime, a Pop III system that also hosts very massive Pop III stars with $M\sim 140$-–260 $M_\odot$ (and lifetimes as short as $\approx 2$ Myr for the models considered here) will already have experienced the explosion of pair-instability supernovae \citep[e.g.][]{Heger02}. If the hosting halo has low mass, these energetic explosions might have cleared away much of the remaining gas from the vicinity of the stars, which would lead to very low ionization parameters and thus reduced HeII$\lambda$1640 equivalent widths. For the lowest-mass Pop III haloes, a large fraction of the gas may even be removed from the halo \citep[e.g.][]{Bromm03,Kitayama05,Mead25}. This would preclude the detection of any emission lines associated with the late evolution of these 20 M$_{\odot}$ stars, unless the gas can quickly recollapse (in the \citealt{Magg22b} simulations of $\sim 10^6 M_\odot$ Pop III haloes, recollapse can happen in as little as $\approx 10$ Myr, although the median recollapse time is 50 Myr). Even without supernovae, the initially dense gas around a cluster of Pop III stars in $\lesssim 10^8\ \mathrm{M}_\odot$ haloes may be expelled to large radii and low densities in response to the radiation field from the stars, with long recombination time-scales and significant leakage of ionizing photons into the intergalactic medium as a result \citep[e.g.][]{Sibony22}. Detailed simulations would be required to assess whether rotating $20\ \mathrm{M}_\odot$ stars would realistically be able to produce strong HeII emission lines (at the level of the recently detected \citealt{Maiolino24} Pop III galaxy candidate) late in the lifetimes of these stars, even if evolution were to follow the \citet{Yoon12} models.

\section{Impact on the 21 cm signal from cosmic dawn}
\label{sec:21cm}

The impact of Pop~III stars on the cosmic 21-cm signal is multifold. During the epoch of cosmic dawn, the Lyman-band continuum flux of Pop III stars plays an important role through the Wouthuysen--Field (WF) effect, wherein Lyman-$\alpha$ photons scatter off neutral hydrogen in the intergalactic medium and act to couple the 21-cm spin temperature to the kinetic temperature of the gas \citep{Wouthuysen1952,Field1958}. The scattering of Lyman-line photons also acts to raise the kinetic temperature of the intergalactic medium \citep[e.g.][]{Chuzhoy07,Reis21}. At the same time, LW photons dissociate H$_2$ molecules, suppressing further Pop~III star formation in molecular cooling dark matter minihaloes \citep[e.g.][]{Fialkov13,Visbal14,Mirocha18}. During the epoch of reionization, Pop~III stars that are still able to form ionize their surrounding gas, leading to a suppression in the 21-cm signal \citep[e.g.][]{Liu25}

\begin{figure}
    \centering
    \includegraphics[width=\linewidth]{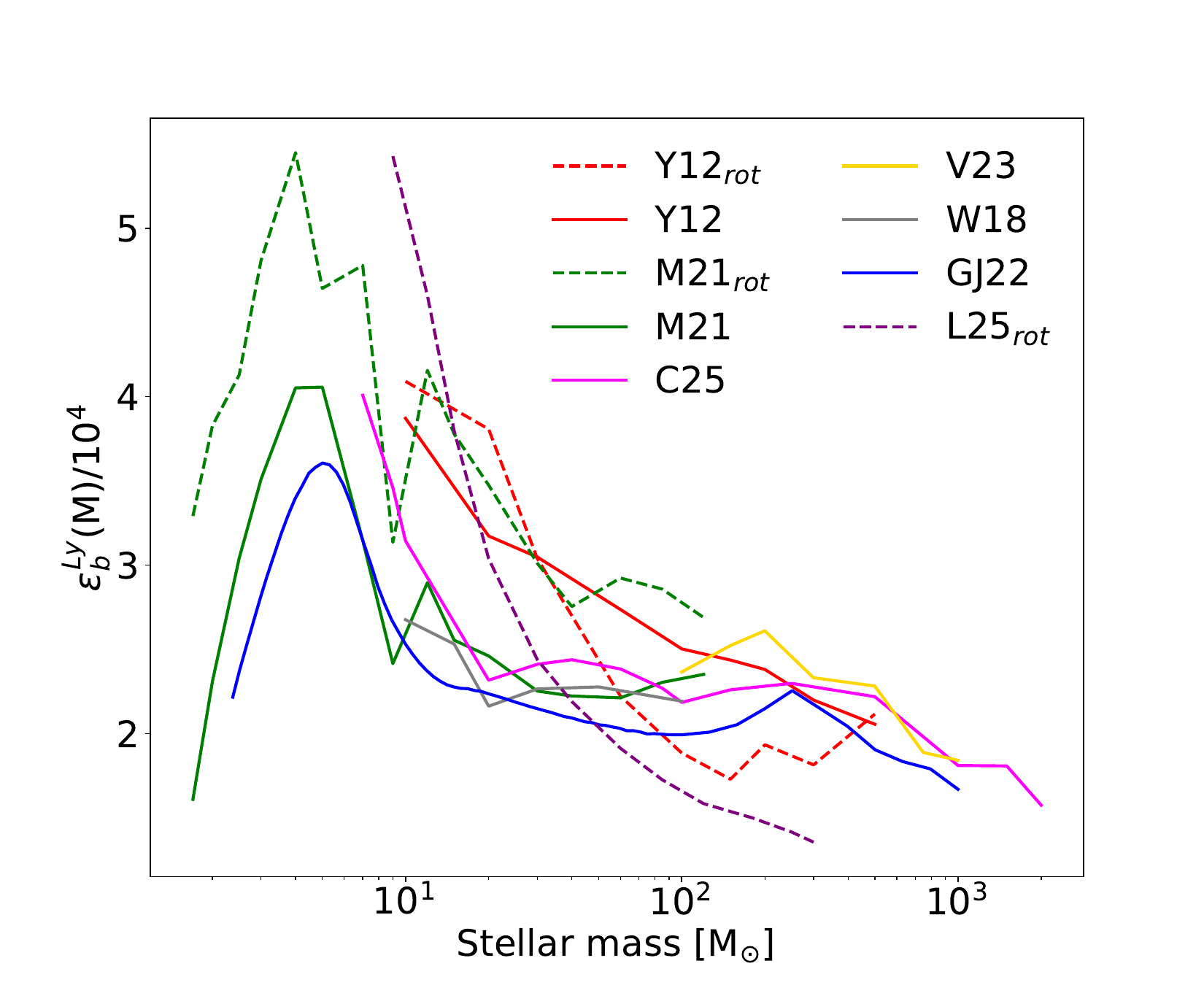}
    \caption{Total Lyman-band emission per stellar baryon as a function of ZAMS mass for the various stellar models considered in this work. Results from \citet[][GJ22]{Gessey-Jones22} are shown in blue for comparison (cf. their fig. 5). The rotating models from \citet[][M21$_\mathrm{rot}$]{Murphy21a} exhibit a clear enhancement in Lyman-band emission, primarily due to their extended lifetimes. In contrast, the \citet[][Y12$_\mathrm{rot}$ and Y12]{Yoon12} and \citet[][L25$_\mathrm{rot}$]{Liu25} models display a more nuanced behaviour: while rotating stars with $M_{\mathrm{ZAMS}} \lesssim 60\,M_{\odot}$ emit more than their non-rotating counterparts from other studies, their emission falls below that of non-rotating \citet{Yoon12} models for $M_{\mathrm{ZAMS}} \gtrsim 30\,M{_\odot}$ and $\gtrsim 20\,M_{\odot}$ respectively.} 
    \label{fig:Lyman_band} 
\end{figure}

In Fig. \ref{fig:Lyman_band}, we plot the lifetime-integrated Lyman-band photon production rates per ZAMS stellar baryon $\epsilon_\mathrm{b}^\mathrm{Ly}$ of our Pop III models in relation to the corresponding predictions presented by \citet{Gessey-Jones22} for non-rotating Pop III stars, and those of \citet{Liu25} for rotating Pop~III stars. Both of these are, like the Muspelheim ones, based on TLUSTY stellar atmosphere SEDs, but employ different assumptions concerning stellar evolution. The \citet{Gessey-Jones22} and \citet{Liu25} models both trace evolution up to the end of main sequence under the assumption of no mass loss, whereas our models also include later evolutionary stages and -- in the case of \citet{Yoon12} models with rotation -- substantial mass loss. Whereas the \citet{Gessey-Jones22} models are based on non-rotating stars, the \citet{Liu25} models are based on \citet{Sibony22} tracks for polytrope stars subject to chemically homogeneous evolution due to strong internal mixing. While \citet{Sibony22} choose to remain agnostic about the physical reason for what would cause this mixing, fast rotation represents a strong candidate, as evident by the rotating \citet{Yoon12} models, which in many cases also undergo chemically homogeneous evolution. For simplicity, we hereafter refer to the \citet{Liu25} models as rotating, in line with the discussion of these models by \citet{Liu25,Liu24}.

Differences in $\epsilon_\mathrm{b}^\mathrm{Ly}$ between the models at any given mass are within about a factor of two, with the rotating stars below 30 M$_{\odot}$ consistently resulting in higher rates compared with their non-rotating counterparts due to their increased lifetimes. Interestingly for our discussion, we note a large difference compared to the \citet{Gessey-Jones22} models at a mass of 20 M$_{\odot}$, where the \citet{Yoon12} models for rotating stars (which have the potential to boost the HeII$\lambda$1640 emission line equivalent width; Fig.~\ref{fig:EW_new}) give rise to an $\epsilon_\mathrm{b}^\mathrm{Ly}$ that is higher by a factor of $\approx 1.7$. Since it is important to establish whether differences of this amplitude could have any significant impact on 21-cm signatures from the pre-reionization Universe, we use the semi-numerical code \textsc{21cmSPACE} \citep[e.g.][]{Fialkov14b,Reis21,Gessey-Jones22,Gessey-Jones25}\footnote{\url{https://www.cosmicdawnlab.com/21cmSPACE}} to simulate the global 21-cm signal and the spherically averaged 21-cm power-spectrum $\Delta_{21}(k,z)$ at $z=6-40$, under the extreme assumption of a single-star IMF where all Pop III stars have the same mass (a rough approximation of the case where the IMF is strongly peaked at this mass). 

The simulations are performed on a cosmological box size of $(384\,\text{cMpc})^3$, and include effects such as baryon dark-matter streaming velocity \citep{Fialkov12}, Lyman--Werner feedback \citep[important during the cosmic dawn;][]{Fialkov13,Munoz22}, Lyman-$\alpha$ heating and scattering \citep{Reis21}, photoheating feedback \citep{Sobacchi13,Cohen16}, X-ray heating \citep{Fialkov14b,Fialkov14c} and ionization from flexible spectra models (important during the epoch of reionization).
In particular, for this work, we use the code version described in \citet{Liu25} where the effect of ionizing photons from Pop~III stars was introduced. To put the \citet{Yoon12} Pop III models into context, we also run \textsc{21cmSPACE} with the Pop~III models of \citet{Gessey-Jones22} and \citet{Liu25}.

We consider two models, with varying Pop~III star formation efficiencies $f_{\star,\mathrm{III}}$ (i.e. fraction of gas converted in stars): \textit{Model 1} where Pop~II SFR dominates at $z \approx 15$ with $f_{\star,\mathrm{III}}=0.1\%$ and \textit{Model 2} where Pop~III SFR is comparable to Pop~II at $z\approx 15$ with $f_{\star,\mathrm{III}}=1\%$, as shown in Figure~\ref{fig:SFRD}. In both models, we fix the Pop~II star-formation efficiency to $f_{\star,\mathrm{II}}=1\%$ which generally matches observations at UV luminosity function observations at $z \lesssim 10$ \citep{Tacchella18,Sipple24,Dhandha25}, and an star-formation averaging timescale of $t_{\star,\mathrm{II}}=0.2H(z)^{-1}$. We also fix $t_\text{rec}=100$ Myr, which is the recovery time for Pop~II star formation after the onset of Pop III supernovae \citep{Magg22a}. To pinpoint the role of Lyman-band emissivities of Pop III stars in shaping the 21-cm signal, we switch off X-ray heating from Pop III stars. For Pop II, we adopt an X-ray luminosity $L_\mathrm{X}$ tied to the star formation rate (SFR) of each halo, according to $L_\mathrm{X}/\mathrm{SFR}=3\times 10^{40}$ erg s$^{-1}$ M$_{\odot}^{-1}$ yr$^{-1}$ \citep{Fragos13a,Fragos13b}, which is consistent with constraints on the diffuse X-ray background, and HERA constraints on the 21-cm power spectrum at $z\lesssim 10$ \citep{Dhandha25}. Throughout, we assume instantaneous emission from Pop~III stars, which is a reasonable assumption for stellar masses $\gtrsim10 \ \mathrm{M}_\odot$ \citep{Gessey-Jones22}.

\begin{figure*}
    \centering
    \includegraphics[width=0.9\linewidth]{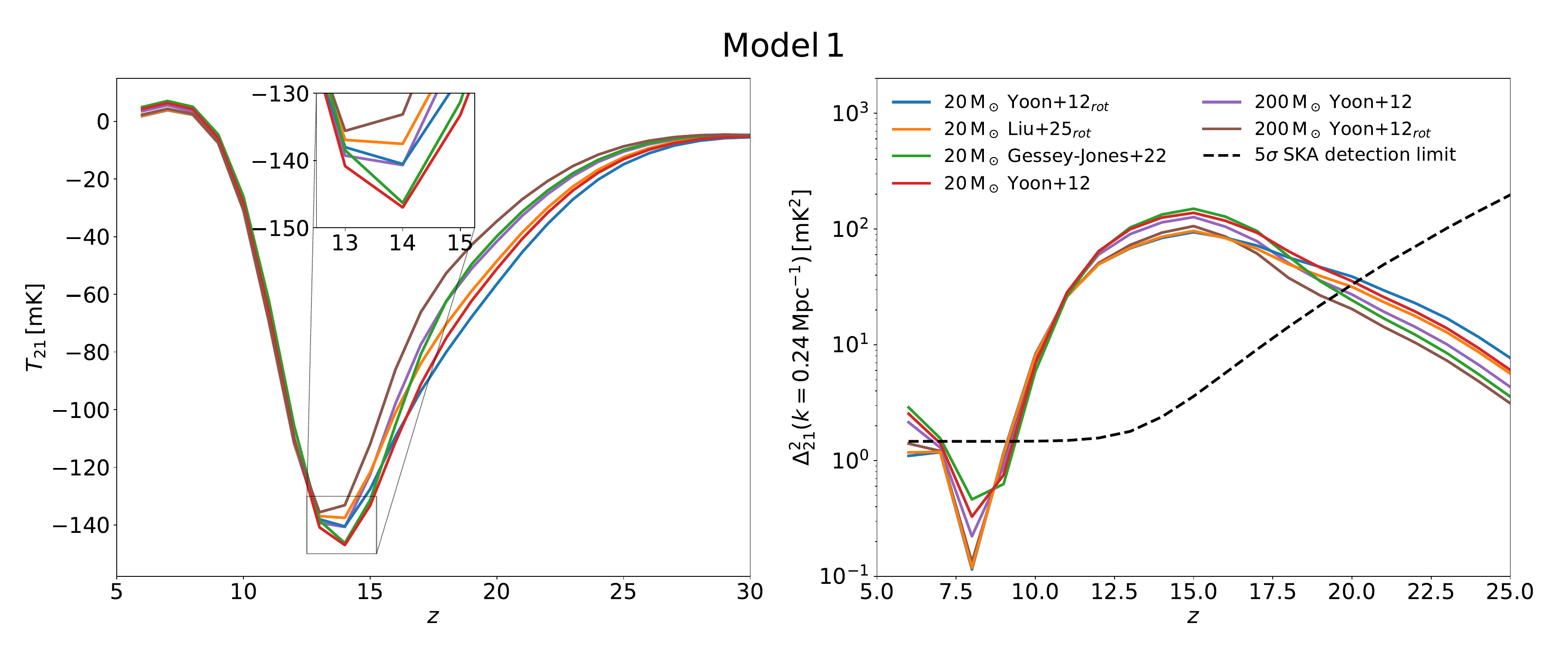}
    \includegraphics[width=0.9\linewidth]{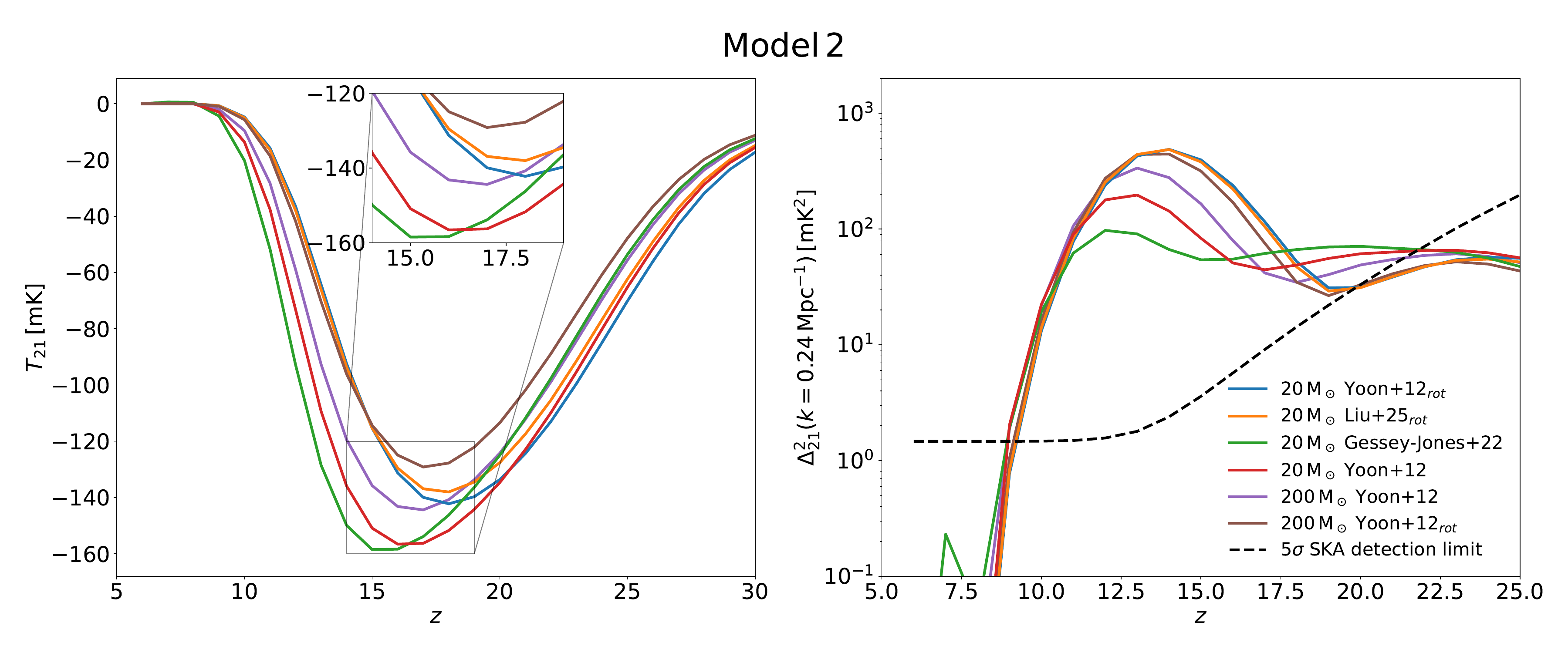}
    \caption{Simulated 21-cm global signal and power spectrum at k = 0.24cMpc$^{-1}$, based on single-mass IMFs ($20$ M$_{\odot}$ or 200 M$_{\odot}$, as specified in the legend) with stellar SEDs from \citet{Yoon12}, \citet{Gessey-Jones22}, and \citet{Liu25}. Model 1 (upper panels) and Model 2 (lower panels) differ in the assumed Pop~III star-formation efficiency ($0.1\%$ and $1\%$ respectively) as discussed in Section~\ref{sec:21cm}. The black line shows the estimated 5$\sigma$ SKA detection limit from a 1080h observation using the array assembly AA* configuration of SKA1-Low within a radius of 2 km from the array center. Differences between the different SED models are very small for both the global 21-cm signal and the 21-cm power spectrum in Model~1, but observationally significant in Model 2. In Model~2, the 21-cm signatures of the 20 M$_{\odot}$ \citet{Yoon12} rotating model (capable of producing very strong HeII$\lambda$1640 emission) deviates from those of the 200 M$_{\odot}$ \citet{Yoon12} models (also able to produce strong HeII$\lambda$1640 emission), which indicates that 21-cm observations may be able to help break degeneracies between such cases. Similarly, the rotating and non-rotating 200 M$_{\odot}$ \citet{Yoon12} models differ in both models demonstrating the potential for the 21-cm signal in differentiating Pop~III stellar properties.}
    \label{fig:21cm_new}
\end{figure*}

Fig. \ref{fig:21cm_new} shows the resulting global 21-cm signal, and the power spectrum generated at a scale of $k=0.24$ cMpc$^{-1}$, for six models: four models assuming a single-mass IMF of 20 M$_{\odot}$ for the rotating and non-rotating \cite{Yoon12} model, non-rotating \citet{Gessey-Jones22} model, and a rotating \cite{Liu25} model, as well as two additional models of single-mass IMFs at 200 M$_{\odot}$ for rotating and non-rotating \cite{Yoon12} stars.
The 200 M$_{\odot}$ non-rotating \cite{Yoon12} model is included here as an example of a type of star that, because of its high $T_\mathrm{eff}$, could produce strong HeII$\lambda$1640 emission despite being non-rotating (Fig.~\ref{fig:EW_new}).

In the case of the 21-cm power spectra, we also plot the expected 5$\sigma$ detection limit for 1080 hours of tracking-scan observations using the Square Kilometre Array (SKA-Low) with 6 hours of observation per night for 180 days. The derived detection limit considers both the thermal noise and the sample variance. To estimate the uncertainty in the power spectrum due to thermal noise, we used the package \texttt{21cmSense\footnote{\url{https://github.com/rasg-affiliates/21cmSense/tree/main}}} \citep{Pober2013, Pober2014}. We only considered the SKA-Low AA$^{*}$  array configuration \citep{sridhar2024} within a radius of 2~km from the array center and assumed the moderate foreground model in \texttt{21cmSense}. 
The sample variance is calculated according to the equation
\begin{equation}
    \sigma_{\mathrm{sample}}^2 = \frac{\sum n_i(k_i-\bar{k})^2}{N-1},
\end{equation}
where N is the total number of counts and $n_i$ is the number of counts in the $k_i$-bin. The total noise is then estimated as the sum of thermal noise and sample variance in quadrature:
\begin{equation}
    \sigma_{\mathrm{total}} =  \sqrt{\sigma_\mathrm{sample}^2 + \sigma_\mathrm{thermal}^2}.
\end{equation}

\subsection{The global 21-cm signal}

The differences in the global signal between the Pop~III stellar models can be explained by considering the epochs $z\sim25$ where the early onset of the global signal absorption trough is almost entirely dictated by the Lyman band emissivities (Fig.~\ref{fig:Lyman_band})\footnote{Although Lyman--Werner photons do contribute to suppression of star-formation in molecular cooling haloes, and we account for this.}, while the absorption trough at $z\sim15$ is also sensitive to the ionizing flux from the stars (Fig.~\ref{fig:Q_lifetime_total}).

During the cosmic dawn ($z\approx20-30$), a higher Lyman-band flux $\epsilon_\mathrm{b}^\mathrm{Ly}$ leads to a stronger WF effect and more rapid onset of significant 21-cm absorption. This would be strongest for the rotating 20 M$_{\odot}$ \citet{Yoon12} model and weakest for the rotating 200 M$_{\odot}$ \citet{Yoon12} model, which indicates that the 21-cm signal observations can have the potential to distinguish the IMF of stellar populations \citep[e.g. as also demonstrated in][]{Gessey-Jones25}. The signal also shows differentiating features between the rotating and non-rotating 200 M$_{\odot}$ \citet{Yoon12} models, and a similar (but reversed) trend for the 20 M$_{\odot}$ \citet{Yoon12} models. In case of the rotating 20 M$_{\odot}$ \citet{Liu25} model, since it exhibits a similar Lyman flux to the non-rotating 20 M$_{\odot}$ \citet{Yoon12}, it shows similar behaviour at high redshifts. The difference in the 21-cm signal is of order few mK at the onset of the absorption signal up to $20\,\text{mK}$ ($\sim20-30\%$) when Lyman-coupling is efficient (at $z \approx 20$ for Model~1 and $z \approx 25$ for Model~2).

During the epoch of heating and reionization ($z\lesssim20$), the differences in these models is dominated by the ionization from Pop~III stars \citep[as explored in][]{Liu25}. The case of rotating models show convergence since they predict similarly strong ionizing fluxes across stellar masses (at $\gtrsim 10$ M$_\odot$), manifested as a suppressed absorption through in both Model~1 and Model~2. In the latter case, the differences are particularly large with a difference also seen in the redshift of the absorption signal as ionization becomes efficient. Indeed, this is reflected in the neutral hydrogen fraction shown in Fig.~\ref{fig:xHI} with a $10-20 \%$ reduction between the non-rotating and rotating cases (and thus, a proportional suppression of the signal). With the non-rotating models, the 20 M$_{\odot}$ \citet{Gessey-Jones22} model has the least efficient ionization, followed by 20 M$_{\odot}$ \citet{Yoon12} model (which shows similar absorption amplitude/timing), and lastly the 200. M$_{\odot}$ \citet{Yoon12} model.

\subsection{The 21-cm power spectrum}

The predicted high-redshift ($z \gtrsim 20$) 21-cm power spectra for Model 1 falls below the $5\sigma$ detection limit of the SKA, and close to the limit for Model 2. At lower redshifts ($z \lesssim 20$), where the power spectrum is deemed detectable according to our error model, the differences between stellar models are marginal in Model~1 (especially if considering degeneracies with other astrophysical parameters that we keep fixed here). However, there are more pronounced and detectable differences in Model 2 which come from the ionization signature of the rotating Pop~III stars, as in the global signal. In these cases, as the Universe is rapidly ionized, the ionized bubbles imprint boosted fluctuations in the power spectrum at $z\approx12.5$. The non-rotating cases show a similar trend to the global signals, where the 20 M$_{\odot}$ \citet{Gessey-Jones22} model shows the smallest ionization peak of $\approx 100$ mK$^2$, while the 20 M$_{\odot}$ \citet{Yoon12} model and 200 M$_{\odot}$ \citet{Yoon12} model show higher peaks by $\times 2$ and $\times 3$. 

Overall, we conclude that the dramatic difference seen between the He$^+$-ionizing rates (Fig. \ref{fig:QHeII_vs_t}) and HeII$\lambda$1640 equivalent widths (Fig. \ref{fig:EW_new}) can imprint some detectable 21-cm signatures. The differences are modest for low Pop~III star-formation efficiency (Model 1), potentially only detectable with high-accuracy global signal experiments (noise levels of $\sim 10$ mK) at $z \approx 20$. For high Pop~III star-formation efficiency (Model 2, which shows early reionization), the differences imprinted in the 21-cm power spectrum at $z \lesssim 20$ are more dramatic and detectable by SKA.

\section{Discussion}
\label{sec:discussion}
\subsection{Blackbody versus stellar atmosphere SEDs}
When comparing our results with the models based on blackbody spectra presented by \citet{Klessen23} in Fig. \ref{fig:Q_lifetime_total}, we find a large discrepancy is in the He$^{+}$ ionizing rates for stars $<80\ \mathrm{M}_{\odot}$, where blackbody spectra result in rates orders of magnitude higher than those based on stellar atmosphere spectra. These differences, also echoed in our \citet{Murphy21a} comparison of momentary $Q_\mathrm{He^{+}}$ for stellar atmosphere spectra versus blackbody spectra in Fig. \ref{fig:bbcomp}, are due to the different shapes of blackbody and stellar atmosphere SEDs and the lack of continuum breaks in the former case, as shown in Fig. \ref{fig:atmos vs BB}. Hence, blackbody spectra should not be used to estimate Pop III photon production rates or fluxes in the He$^{+}$ ionizing energy range. 

We also note significant deviations in the other UV bands for stars $<9\ \mathrm{M}_{\odot}$. %when comparing blackbody and stellar atmospheres, as shown in Fig. \ref{fig:Q_lifetime_total}. 
While \citet{Murphy21b} did comment on the reliability of the blackbody approximation for the \citet{Murphy21b} tracks (through a comparison to the results of \citealt{Schaerer02}), this offset was not noted since the \citet{Murphy21b} study of UV photon production rates considered only stars of masses $\geq9\ \mathrm{M}_{\odot}$ from the \citet{Murphy21a} set. The blackbody approximation from their lower-mass tracks is adopted by \citet{Klessen23} which overestimates the ionizing rates for all low mass stars and slightly underestimates the LW rates for these same stars. However, as long as the Pop III IMF has a characteristic mass $\gtrsim9\ \mathrm{M_{\odot}}$ -- which is often argued to be the case \citep[e.g.][]{Hartwig24,Liu24b} -- this discrepancy should not be a major source of error, as the ionizing photon production in that case would be dominated by the more massive stars. 

\subsection{The Lyman--Werner emissivities of Pop III stars}
\label{subsec: LW}
The LW photon production rates of Pop III stars are important for establishing the LW radiation background during cosmic dawn and for the quenching of subsequent star formation in low-mass haloes. When the lifetime-integrated $Q_\mathrm{LW}$ is expressed as LW photons per stellar baryon, there are two distinct values that are frequently featured in the literature:  4800 and  $\sim 10^5$. Our models do not favour either of these, but instead favour values in the intermediate range $\approx 8\times 10^3$ to $3\times 10^4$ (see Fig. \ref{fig:Q_lifetime_total}; $\log Q_\mathrm{LW}/M_\mathrm{ZAMS}$ values in this figure can be converted to log LW photons per stellar baryon by subtracting $\approx 57.08$).
The lower literature value (4800) stems from \citet{Barkana05}, based on Pop III spectra from \citet{Bromm01} and an adopted power-law slope of the spectrum, whereas the higher value ($\sim 10^5$) is usually attributed to \citet{Schaerer02}. Indeed, when the \citet{Schaerer02} Table 4 values are converted into these units, they peak at $\sim 9\times 10^4$ LW photons per baryon at a Pop III stellar mass of 60 M$_{\odot}$. However, as shown in Fig. ~\ref{fig:Q_lifetime_total}, the \citet{Schaerer02} $Q_\mathrm{LW}$ are higher than ours at $M\gtrsim 10\ \mathrm{M}_\odot$ by up to a factor $\approx 5$ (0.7 dex). \citet{Mas-Ribas16} also note difficulties in reproducing the \citet{Schaerer02} LW rates and suggest that the \citet{Schaerer02} LW rates may have been integrated from 11.2 eV (wavelength 1107 \AA) to infinity, rather than to 13.6 eV (wavelength 912 \AA), which would lead to boosted rates. Lifetime-integrated LW rates per stellar baryons in the $\sim 10^5$ range for Pop III stars are also reported in the recent paper by \citet{Costa25}, and in that case the LW rates were indeed integrated from 11.2 eV to infinity (Guglielmo Costa, private communication). After correcting the \citet{Costa25} limit from 11.2 to 13.6 eV, their values agree with ours. Our values are moreover in good agreement with those recently presented by \citet{Liu25} for the non-rotating Pop III tracks of \citet{Gessey-Jones22} and the rotating Pop III tracks of \citet{Sibony22}. Hence, our recommendation is to use values for the lifetime-integrated LW photon counts per stellar baryons of $\approx 8\times 10^3$ to $3\times 10^4$. 

\subsection{Viability of the stellar atmosphere models}
The results in this paper are based on SED models for Pop III stars, which at $T_\mathrm{eff}>15000$ K make use of TLUSTY \citep{Hubeny95} stellar atmosphere models for plane-parallel, hydrostatic stars with primordial surface composition. While certainly more realistic than blackbody SEDs, such stellar atmosphere models may still represent an oversimplification that could impact the emitted ionizing photon production rates of Pop III stars. The surface He/H composition of Pop III stars may change over the course of their lifetimes, and the surface may also get polluted by metals produced within the star, due to either chemically homogeneous evolution for a rapidly rotating star \citep[e.g.][]{Yoon12,Liu25}, but also due to dredge-ups for a non-rotating star \citep[e.g.][]{Volpato23}. 

While \citet{Schaerer02} notes that the impact of the H/He level on the stellar atmosphere spectra is small, this statement likely does not apply to models for rotating stars with helium mass fractions close to unity \citep{Sibony22}. \citet{Liu25} do take the evolving helium abundance into account in their modelling of Pop III stars, and while they note that this helps the stellar atmosphere model converge, no analysis of the impact of the pure-helium atmosphere on the emerging stellar atmosphere spectra was presented in their paper.

Indeed, as shown in \citet{Zackrisson24}, the set of TLUSTY stellar atmosphere models used by Muspelheim (with fixed H/He fractions) are -- due to convergence problems -- sometimes unable to reach the low surface gravities of all Pop III stellar evolutionary tracks. This is for instance the case of the $T_\mathrm{eff}\sim 2\times 10^5$ K state of the rotating 20 M$_{\odot}$ \citet{Yoon12} stars highlighted in Section \ref{sec:HeII1640} in relation to its potential to produce a very strong HeII$\lambda$1640 emission line. These 20 M$_{\odot}$ rotating stars lie $\Delta\log(g)\approx 0.25$ below the last converged grid data point. However, based on how the He$^{+}$-flux changes with $\log(g)$ within the converged grid at these $T_\mathrm{eff}$, this is likely to have a very minor effect on $Q_{\mathrm{He}^{+}}$.

Rotational mixing and surface composition can also be affected by the presence of a companion Pop III star in a binary system \citep{Song20}. Rapid rotation could furthermore drive a mechanical wind, and metals at the surface of Pop III stars may also trigger line-driven winds \citep[e.g.][]{Liu21,Jeena23}. The effects of winds on the emerging ionizing photon production rates of Pop III stars were explored by \citet{Schaerer02}, who concluded that the ionizing photon rates would not be substantially affected compared to TLUSTY stellar atmosphere spectra in their models, even in their high mass-loss case. However, \citet{Schaerer02} also notes that for sufficiently dense winds, the He$^{+}$ flux could potentially get substantially suppressed (as demonstrated by \citealt{Chen15} for metal-enriched stars). It may therefore be interesting to revisit this issue in the future, using stellar atmosphere spectra with updated wind models for Pop III stars. However, it should be noted that if winds are launched only late in the evolution of Pop III stars, this is unlikely to matter for the time-integrated emission of ionizing photons into the surrounding medium, unless the Pop III stars also exhibit high $T_\mathrm{eff}$ in such stages.

Finally, rapidly rotating stars would not be able to retain spherical shapes, and would exhibit different $T_\mathrm{eff}$ and surface gravities at their poles compared to their equators, thereby exhibiting anisotropic SEDs. While orientations would be randomized in a large population of such stars, mechanical mass-loss due to rotation may also preferentially happen in the direction perpendicular to the rotation axis, which could introduce a systematic bias in the UV emissivities seen by the surrounding interstellar and intergalactic medium, compared with the simplified models used here. 

\subsection{Binary Pop III stars}
In this paper, we have exclusively considered models for single stars, even though Pop III stars in simulations often form binary and higher-order systems \citep[e.g.][]{Stacy13,Sugimura20}. As shown by \citet{Tsai23}, binary interactions can boost the UV fluxes of Pop III stars (in the case of lifetime-integrated fluxes by up to a factor of $\approx 2$ in the LW band, which is the strongest effect seen in their models). %In some cases, binary interaction can also produce extremely hot stars at high ages \citet{Song20}, similar to the rotating Pop III stars of \citet{Yoon12} studied in this paper.
However, binary interactions are also expected to affect stellar rotation rates \citep[e.g.][]{deMink13}, which by themselves affect stellar evolution and consequently these fluxes. Given the many open questions regarding the stellar initial mass function, initial rotation rate and binary interactions of Pop III stars, models of the UV flux from these stars are likely to suffer from significant uncertainties until the arrival of better observational constraints. As rotation and binary interactions affect the end states of Pop III stars, such constraints may for instance come from the study of Pop III supernova yields in second generation stars or gas absorption systems, from direct observations of Pop III supernovae or gamma-ray bursts (and their afterglows), or from future gravitational wave detectors.

\section{Conclusions}
\label{sec:conclusion}
In this work, we have presented a set of publicly available models for the UV photon production rates of single Pop III stars, with and without rotation, across a wide range of stellar masses. We have investigated how the photon production rates are affected by the adoption of blackbody spectra compared with stellar atmosphere spectra. We have also investigated and compared the resulting lifetime-integrated emissivities of several different sets of Pop III tracks. Lastly, for selected models, we have explored the potential role of stellar rotation on the equivalent width of the HeII$\lambda$1640 emission line, on the 21-cm global signal, and on the the spherically averaged 21-cm power spectrum. 

Our conclusions can be summarized as follows:
\begin{enumerate}
\item The assumption of blackbody spectra generally overestimate He$^{+}$ ionizing rates, in some cases by more than an order of magnitude, depending on the $T_\mathrm{eff}$ of the star. 
\item Based on the stellar evolutionary tracks of \citet{Yoon12}, which take rotation and magnetic fields into account and predict evolution to $T_\mathrm{eff}>10^5$ K at the end stages of evolution, we show that rotation significantly increases the total He$^{+}$ ionizing photon production of Pop III stars in the range 20 to 100 M$_{\odot}$ compared with stars without rotation.
\item If gas can be retained with high-redshift Pop III systems dominated by rotating $\sim 20\ \mathrm{M}_\odot$ stars, then such systems may be able to produce very strong HeII$\lambda$1640 emission with rest-frame equivalent widths comparable to or higher than systems dominated by $\gtrsim 100\ \mathrm{M}_\odot$ Pop III stars. 
\item Differences between rotating and non-rotating Pop III stars in 21-cm observables like the 21-cm global signal and the 21-cm power spectrum are modest for sub-per cent level Pop~III star-formation efficiency, and more pronounced for per cent Pop~III star-formation efficiency, with potential for detectability with the SKA. The differences are of order $\sim 20$ per cent for the global signal (at $z \approx 15$) and a factor of $2-3$ for the ionization peak in the power spectrum (at $z \approx 12.5$) due to the extreme ionizing flux of rotating stars.
\item We highlight that two disparate values for the Lyman--Werner rates of Pop III stars have been used in the previous literature on 21-cm cosmology and high-redshift star formation ($\sim 5\times 10^3$ vs $\sim 1\times 10^5$ LW photons per stellar baryon). Our models instead favour a value around $\sim 10^4$ LW photons per stellar baryon, in line with other recent estimates \citep[e.g.][]{Liu25}.
\end{enumerate}

\section*{Acknowledgements}
This manuscript is based on work initially conducted as part of the first author’s Master’s thesis \citet{Wasserman25}. The work has been expanded upon and revised in collaboration with the co-authors for this publication. EZ acknowledges project grant 2022-03804 from the Swedish Research Council (Vetenskapsr\aa{}det). This work was supported in part by the International Centre for Theoretical Sciences (ICTS) for participating in the program -  Radio Cosmology and Continuum Observations in the SKA Era: A Synergic View. (code: ICTS/radiocoscon2025/04). An anonymous referee provided useful comments that helped improve the quality of the manuscript.
%%%%%%%%%%%%%%%%%%%%%%%%%%%%%%%%%%%%%%%%%%%%%%%%%%
\section*{Data Availability}
\label{sec:data availability}
Publicly available grids of  Muspelheim models (SEDs, $Q(t)$) and both lifetime-integrated and lifetime-average photon production rates are available from: \url{https://www.astro.uu.se/~ez/muspelheim/muspelheim.html}
% time-integrated $Q$



\bibliographystyle{mnras}
\bibliography{references} % if your bibtex file is called example.bib

%\begin{thebibliography}{99}
%\end{thebibliography}

%%%%%%%%%%%%%%%%%%%%%%%%%%%%%%%%%%%%%%%%%%%%%%%%%%

%%%%%%%%%%%%%%%%% APPENDICES %%%%%%%%%%%%%%%%%%%%%

\appendix

\section{Extra tables and figures}
\label{sec:appendix}

Table~\ref{tab:Qs} shows the average lifetime photon production rates for the \citet{Yoon12} and \citet{Murphy21a} models used in this work. Fig.~\ref{fig:xHI} shows the neutral fractions from the \textsc{21cmSPACE}\ simulations in Fig.~\ref{fig:21cm_new}.

\begin{table*}
\caption{Average lifetime photon production rates. The first column is the model source, the second gives the initial rotational velocity in fraction of the Keplerian velocity for \citet{Yoon12} and critical velocity for \citet{Murphy21a}. The third column is the ZAMS mass in solar masses, the fourth through eighth column is the logarithm of the average lifetime photon production rates [$s^{-1}$] for the respective energy ranges, and the last column is the lifetimes in years.}
\label{tab:Qs}
\begin{tabular}{ccccccccc}
\hline
Source   & Rotation & Mass & $\overline{Q}_{H}$ & $\overline{Q}_{He}$ & $\overline{Q}_{He^{+}}$ & $\overline{Q}_{LW}$ & $\overline{Q}_{Ly}$ & Lifetime\\ \hline
Yoon+12   & 0.0      & 10   & 47.71    & 47.23     & 41.98      & 47.57    & 47.75    & 2.61E+07 \\
Yoon+12   & 0.0      & 20   & 48.66    & 48.31     & 43.79      & 48.20    & 48.37    & 1.02E+07 \\
Yoon+12   & 0.0      & 30   & 49.10    & 48.77     & 44.87      & 48.54    & 48.70    & 6.81E+06 \\
Yoon+12   & 0.0      & 60   & 49.68    & 49.39     & 47.66      & 49.00    & 49.17    & 4.16E+06 \\
Yoon+12   & 0.0      & 100  & 50.02    & 49.74     & 48.54      & 49.29    & 49.46    & 3.27E+06 \\
Yoon+12   & 0.0      & 150  & 50.27    & 50.00     & 48.98      & 49.54    & 49.71    & 2.70E+06 \\
Yoon+12   & 0.0      & 200  & 50.40    & 50.14     & 49.19      & 49.66    & 49.83    & 2.64E+06 \\
Yoon+12   & 0.0      & 300  & 50.59    & 50.33     & 49.45      & 49.85    & 50.01    & 2.40E+06 \\
Yoon+12   & 0.0      & 500  & 50.83    & 50.57     & 49.73      & 50.08    & 50.25    & 2.18E+06 \\
Yoon+12   & 0.4      & 10   & 47.56    & 46.98     & 41.53      & 47.57    & 47.75    & 2.74E+07 \\
Yoon+12   & 0.4      & 20   & 48.88    & 48.63     & 47.48      & 48.12    & 48.28    & 1.52E+07 \\
Yoon+12   & 0.4      & 30   & 49.26    & 49.03     & 48.00      & 48.40    & 48.56    & 9.51E+06 \\
Yoon+12   & 0.4      & 60   & 49.79    & 49.59     & 48.75      & 48.82    & 48.98    & 5.28E+06 \\
Yoon+12   & 0.4      & 100  & 50.12    & 49.94     & 49.21      & 49.10    & 49.26    & 3.93E+06 \\
Yoon+12   & 0.4      & 150  & 50.36    & 50.18     & 49.48      & 49.32    & 49.47    & 3.28E+06 \\
Yoon+12   & 0.4      & 200  & 50.52    & 50.32     & 49.59      & 49.54    & 49.69    & 2.97E+06 \\
Yoon+12   & 0.4      & 300  & 50.69    & 50.48     & 49.68      & 49.74    & 49.90    & 2.55E+06 \\
Yoon+12   & 0.4      & 500  & 50.87    & 50.62     & 49.74      & 50.07    & 50.25    & 2.23E+06 \\
Murphy+21 & 0.0      & 9    & 47.68    & 47.13     & 43.98      & 47.30    & 47.61    & 2.00E+07 \\
Murphy+21 & 0.0      & 12   & 48.10    & 47.63     & 44.86      & 47.48    & 47.83    & 1.93E+07 \\
Murphy+21 & 0.0      & 15   & 48.23    & 47.90     & 45.47      & 47.55    & 48.01    & 1.40E+07 \\
Murphy+21 & 0.0      & 20   & 48.84    & 48.40     & 46.19      & 48.07    & 48.26    & 1.03E+07 \\
Murphy+21 & 0.0      & 30   & 49.07    & 48.82     & 46.95      & 47.91    & 48.58    & 6.64E+06 \\
Murphy+21 & 0.0      & 40   & 49.45    & 49.16     & 47.40      & 47.87    & 48.81    & 5.21E+06 \\
Murphy+21 & 0.0      & 60   & 49.81    & 49.52     & 47.94      & 48.34    & 49.09    & 4.03E+06 \\
Murphy+21 & 0.0      & 85   & 50.15    & 49.79     & 48.37      & 49.31    & 49.34    & 3.40E+06 \\
Murphy+21 & 0.0      & 120  & 50.38    & 50.02     & 48.72      & 49.60    & 49.54    & 3.05E+06 \\
Murphy+21 & 0.4      & 9    & 47.57    & 47.06     & 41.73      & 47.47    & 47.65    & 2.38E+07 \\
Murphy+21 & 0.4      & 12   & 47.93    & 47.52     & 42.42      & 47.74    & 47.92    & 2.25E+07 \\
Murphy+21 & 0.4      & 15   & 48.26    & 47.89     & 43.02      & 47.94    & 48.11    & 1.65E+07 \\
Murphy+21 & 0.4      & 20   & 48.64    & 48.30     & 43.73      & 48.18    & 48.34    & 1.19E+07 \\
Murphy+21 & 0.4      & 30   & 49.09    & 48.77     & 44.77      & 48.49    & 48.65    & 7.60E+06 \\
Murphy+21 & 0.4      & 40   & 49.36    & 49.05     & 45.92      & 48.70    & 48.86    & 5.76E+06 \\
Murphy+21 & 0.4      & 60   & 49.70    & 49.39     & 47.50      & 49.01    & 49.18    & 4.41E+06 \\
Murphy+21 & 0.4      & 85   & 49.94    & 49.64     & 48.20      & 49.24    & 49.41    & 3.60E+06 \\
Murphy+21 & 0.4      & 120  & 50.17    & 49.89     & 48.65      & 49.43    & 49.59    & 3.14E+06 \\ \hline
\end{tabular}
\end{table*}

\begin{figure*}
    \centering
    \includegraphics[width=1\linewidth]{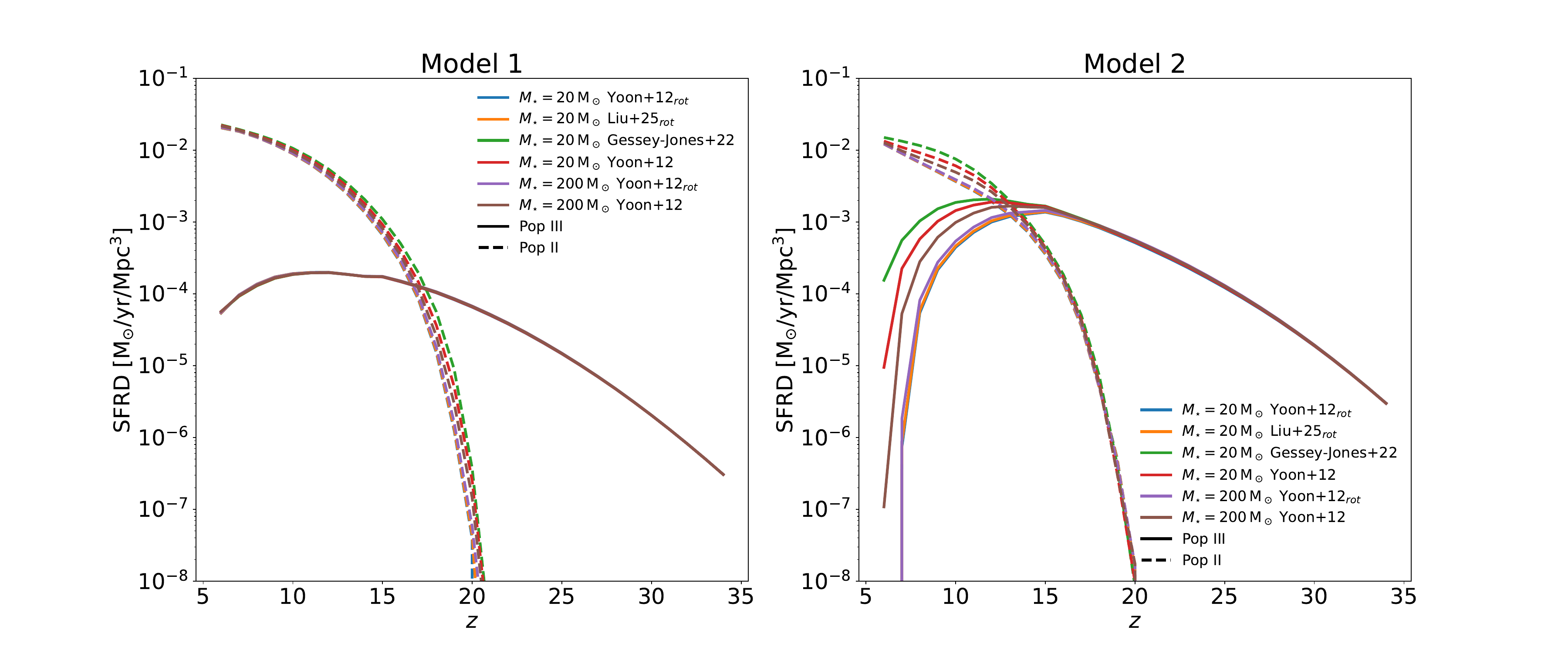}
    \caption{Star formation rate density as a function of redshift for the \textsc{21cmSPACE} simulations models included in Fig. \ref{fig:21cm_new}. } 
    \label{fig:SFRD}
\end{figure*}

\begin{figure*}
    \centering
    \includegraphics[width=0.85\linewidth]{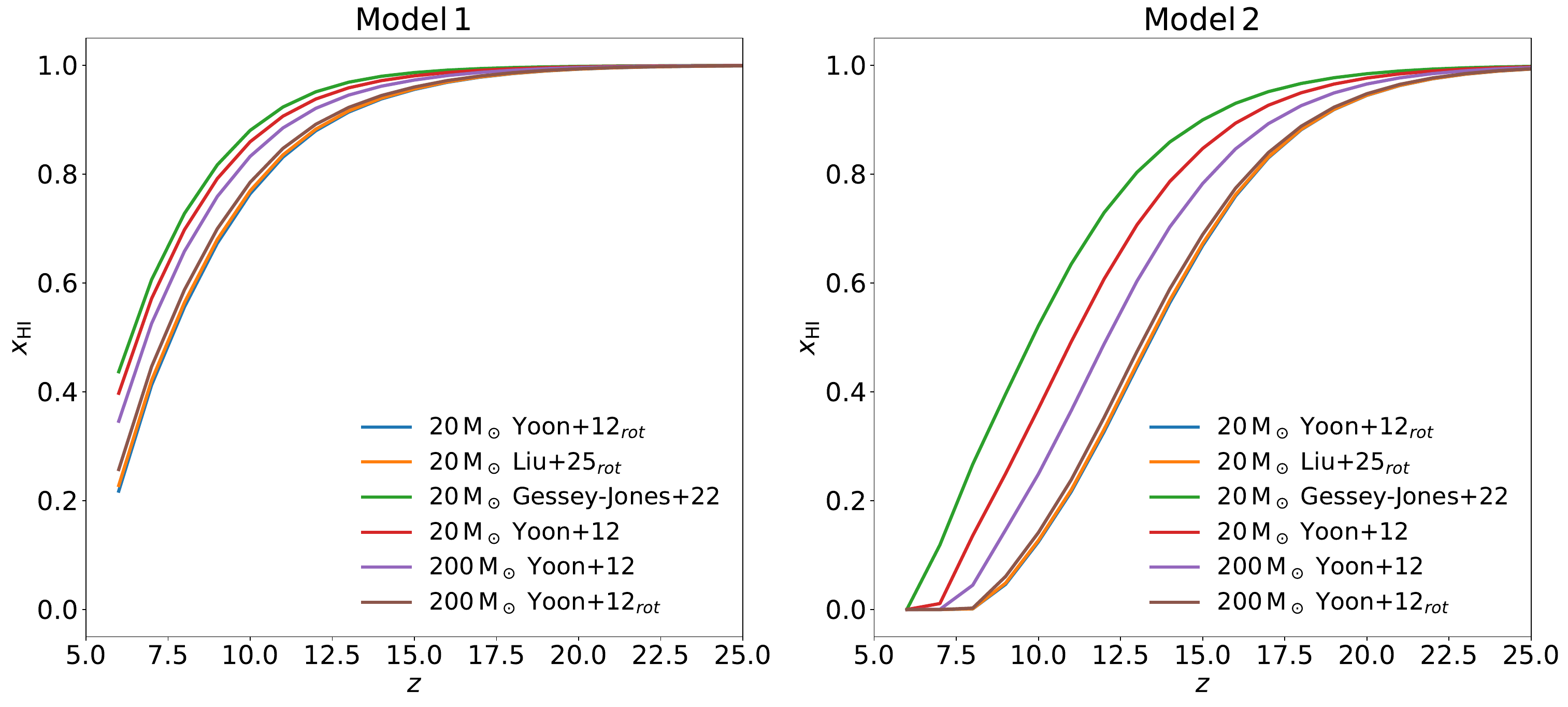}
    \caption{Neutral hydrogen fractions as a function of redshift for the \textsc{21cmSPACE} simulations models included in Fig. \ref{fig:21cm_new}. For the rotational stellar models in scenario 2, the Universe is already ionized to $x_\text{HI}\approx0.8$ by $z\approx15$.} 
    \label{fig:xHI}
\end{figure*}

%%%%%%%%%%%%%%%%%%%%%%%%%%%%%%%%%%%%%%%%%%%%%%%%%%

% Don't change these lines
\bsp	% typesetting comment
\label{lastpage}
\end{document}